\documentclass[
journal=largetwo,
manuscript=research-paper,
year=2020,
volume=37,
]{cup-journal}

\usepackage{amsmath}
\usepackage[nopatch]{microtype}
\usepackage{booktabs}

\usepackage{hyperref}
\hypersetup{
	colorlinks=true,
	linkcolor=blue,
	filecolor=magenta,      
	urlcolor=cyan,
	pdftitle={Overleaf Example},
	pdfpagemode=FullScreen,
	citecolor=blue,
}

\title{Strategy for mitigation of systematics for EoR experiments with the Murchison Widefield Array}
\author{C.~D.~Nunhokee}
\affiliation{International Centre for Radio Astronomy Research (ICRAR), Curtin University, Bentley, WA, Australia}
\alsoaffiliation{ARC Centre of Excellence for All Sky Astrophysics in 3 Dimensions (ASTRO 3D), Bentley, Australia}
\alsoaffiliation{Curtin Institute of Radio Astronomy, GPO Box U1987, Perth, WA 6845, Australia}

\author{D.~Null}
\affiliation{International Centre for Radio Astronomy Research (ICRAR), Curtin University, Bentley, WA, Australia}
\alsoaffiliation{Curtin Institute of Radio Astronomy, GPO Box U1987, Perth, WA 6845, Australia}

\author{C.~M.~Trott}
\affiliation{International Centre for Radio Astronomy Research (ICRAR), Curtin University, Bentley, WA, Australia}
\alsoaffiliation{ARC Centre of Excellence for All Sky Astrophysics in 3 Dimensions (ASTRO 3D), Bentley, Australia}
\alsoaffiliation{Curtin Institute of Radio Astronomy, GPO Box U1987, Perth, WA 6845, Australia}

\author{C.~H.Jordan}
\affiliation{International Centre for Radio Astronomy Research (ICRAR), Curtin University, Bentley, WA, Australia}
\alsoaffiliation{ARC Centre of Excellence for All Sky Astrophysics in 3 Dimensions (ASTRO 3D), Bentley, Australia}
\alsoaffiliation{Curtin Institute of Radio Astronomy, GPO Box U1987, Perth, WA 6845, Australia}

\author{J.~B.~Line}
\affiliation{International Centre for Radio Astronomy Research (ICRAR), Curtin University, Bentley, WA, Australia}
\alsoaffiliation{ARC Centre of Excellence for All Sky Astrophysics in 3 Dimensions (ASTRO 3D), Bentley, Australia}
\alsoaffiliation{Curtin Institute of Radio Astronomy, GPO Box U1987, Perth, WA 6845, Australia}

\author{R.~Wayth}
\affiliation{International Centre for Radio Astronomy Research (ICRAR), Curtin University, Bentley, WA, Australia}
\alsoaffiliation{ARC Centre of Excellence for All Sky Astrophysics in 3 Dimensions (ASTRO 3D), Bentley, Australia}
\alsoaffiliation{Curtin Institute of Radio Astronomy, GPO Box U1987, Perth, WA 6845, Australia}

\author{N.~Barry}
\affiliation{International Centre for Radio Astronomy Research (ICRAR), Curtin University, Bentley, WA, Australia}
\alsoaffiliation{ARC Centre of Excellence for All Sky Astrophysics in 3 Dimensions (ASTRO 3D), Bentley, Australia}
\alsoaffiliation{Curtin Institute of Radio Astronomy, GPO Box U1987, Perth, WA 6845, Australia}
\email[C.~D.~Nunhokee]{ridhima.nunhokee@curtin.edu.au}

 %% First letter not capped

\begin{document}

\begin{abstract}
	Observations of the 21 cm signal face significant challenges due to bright astrophysical foregrounds that are several orders of magnitude higher than the brightness of the hydrogen line, along with various systematics. Successful 21 cm experiments require accurate calibration and foreground mitigation. Errors introduced during the calibration process such as systematics, can disrupt the intrinsic frequency smoothness of the foregrounds, leading to power leakage into the Epoch of Reionisation (EoR) window. Therefore, it is essential to develop strategies to effectively address these challenges. In this work, we adopt a stringent approach to identify and address suspected systematics, including malfunctioning antennas, frequency channels corrupted by radio frequency interference (RFI), and other dominant effects. We implement a statistical framework that utilises various data products from the data processing pipeline to derive specific criteria and filters. These criteria and filters are applied at intermediate stages to mitigate systematic propagation from the early stages of data processing. Our analysis focuses on observations from the Murchison Widefield Array (MWA) Phase I configuration. Out of the observations processed by the pipeline, our approach selects $18\%$, totalling 58 hours, that exhibit fewer systematic effects. The successful selection of observations with reduced systematic dominance enhances our confidence in achieving 21 cm measurements.
	
%	Observations of 21 cm signal are challenged by bright astrophysical foregrounds {\bf several orders of magnitude}  higher than the hydrogen line along with systematics. 21 cm experiments englobe two fundamental requirements: calibration and foreground mitigation. 
%	Errors induced in the calibration process due to systematics corrupt the intrinsic frequency smoothness of the foregrounds causing power to leak into the EoR window. It is therefore vital to place our efforts towards formulating a strategy to deal with them. This work adopts a strict approach to treat any suspected systematics, including malfunctioning antennas, bad timestamps and frequency channels and observations corrupted by RFI or other dominant effects. We implement a statistical framework that builds upon various data products from the data processing pipeline to derive certain criteria and filters that are then imposed at intermediate stages to prevent systematic propagation from the very early stage. Observations from the MWA Phase I configuration are used. Out of the observations that were fed to the pipeline, our approach selected $18\%$, a total of 58~hours. The successful set of observations is regarded as less systematic dominated raising our confidence in the resulting 21 cm measurements.
 \end{abstract}	

\section{Introduction}
The Epoch of Reionisation (EoR) marked a significant transition in the history of the Universe,  about 400 million years after the Big Bang, when the first galaxies along with other cosmic structures formed, and the intergalactic medium transitioned from a neutral state to an ionised state. This epoch unveils a wealth of information including formation and ionisation of the first cosmic structures aiding us to gain insights into the early universe's physical processes. One way to study this era is through mapping of neutral hydrogen and subsequently tracing its evolution. Neutral hydrogen emits and absorbs radiation at the specific wavelength of 21~cm. The 21~cm hydrogen line (HI)  corresponds to the transition between two energy states of the hydrogen atom, specifically the spin-flip transition of the electron in ground state.

The 21~cm HI can be mapped through its spatial fluctuations, measured from the difference in brightness temperatures across the intergalactic medium, encoding information about the density and ionisation state of neutral hydrogen, as well as the clustering and growth of cosmic structures.
The underlying physical processes driving the EoR can be inferred through statistical properties of these fluctuations using power spectrum analysis \citep{Furlanetto2016}. Experiments such as the Giant Metrewave Radio Telescope Epoch of Reionisation \citep[GMRT,][]{Paciga2011}, the Hydrogen Epoch of Reionisation \citep[HERA,][]{DeBoer2017, Berkhout2024}, the Murchison WideField Array \citep[MWA,][]{Tingay2013, Randall2018} and the LOw Frequency ARray \citep[LOFAR,][]{vanHaarlem2013} are currently focused at the statistical detection of the 21~cm HI. The aforementioned instruments are a combination of both first and second-generation telescopes focused at studying large-scale structures before and during reionisation between redshifts of 6 to 11.  HERA has recently reported the most stringent upper limits of $\Delta^2 \leq (21.4 \, \textrm{mK})^2$ at $k=0.34 \, h$ Mpc$^{-1}$ and $\Delta^2 \leq  (59.1 \, \textrm{mK})^2$ at $k=0.36 \,h$Mpc$^{-1}$ at redshifts of $7.9$ and $10.4$ respectively, placing new constraints on astrophysical parameters of reionisation. These results suggest heating of the intergalactic medium above the adiabatic cooling limit must have occurred by at least $z=10.4$ \citep[]{HERACollaboration2022A, HERA2023}.

While 21~cm experiments hold great potential to enhance our understanding on the evolution of the early Universe, they are challenged by strong astrophysical foregrounds, both Galactic and extraGalactic, several orders of magnitude higher than the 21~cm HI. To date, two fundamental approaches have been applied solely or in a hybrid system to mitigate foreground contamination: 1) the subtraction method whereby foregrounds are modelled and subtracted from the data \citep{Mitchell2008,Morales2006, Bernardi2010, Bernardi2013,Morales2019}; and 2) the avoidance scheme where foregrounds are constrained to lower spatial scales and an `EoR window' is defined \citep{Morales2004, Vedantham2012, Liu2014a, Liu2014b}. However, both techniques are prone to calibration errors \citep{Barry2016, Patil2016, Barry2017}, uncertainties in foreground and primary beam \citep[i.e.,][]{Neben2016, Procopio2017, EwallWice2017, Nunhokee2020}, and systematics such as Radio Frequency Interferences (RFI), instrumental polarisation leakage and mutual coupling between antennas, etc \citep{Moore2017, Nunhokee2017, Wilensky2019, Barry2019, Kern2020, Alec2022, Charles2022, Kolopanis2023, Wilensky2023, Murphy2023}. These errors, if not treated can potentially lead to biases and leakages in our EoR measurements. Our work presents a strategy to mitigate these systematics by quantifying them for each observation through a set of metrics to prevent them from propagating to the power spectrum.
%This work uses observations from the MWA. 

The paper is broken down such that Section~\ref{sec:methodology} introduces the methodology followed by details of observation setup in Section~\ref{sec:observations}. The data processing pipeline is discussed in Section~\ref{sec:eor_pipeline} where detailed explanation of each stage is provided. Section~\ref{sec:power_spectrum} describes the power spectrum analysis and section~\ref{sec:data_assurance} talks on the data quality assessment. Results are interpreted in Section~\ref{sec:results} and conclusions are drawn in Section~\ref{sec:conclusion}.

%In this paper we would be working with observations from MWA, describing the systematics present and ways in which we mitigated those effects.
%Breakdown of the paper...

%%%%%%%%%%%%%%%%%%%%%%%%%%%%%%%%%%%
\section{Methodology}
\label{sec:methodology}
 {Astronomers have been grappling with systematic propagation to avoid biases in the 21~cm power spectra measurements for years. Two fundamental methods have been employed to date:}
\begin{enumerate}
	\item Identify potential systematics and discard or flag  them.
	\item Identify potential systematics and apply mitigation techniques to address them.
\end{enumerate}

Systematics can arise from the presence of corrupted timestamps, corrupted frequency channels, RFI sources, instrumental leakages, as well as from unknows origins. Efforts have been dedicated towards detecting the known systematic sources and alleviating them. However, we are not confident about the goodness of data points surrounding the corrupted ones. We could potentially extend the flags to neighbours, but this avenue may turn into an indefinite process \citep{Offringa2010}. A quantitative approach to how much the identified flags could leak into the non-identified is required and this demands precise understanding of the systematic source. Further, mitigation techniques struggle to remove systematics of unknown origins which ultimately leave some traces behind. These residual systematics, even with low intensities could potentially harm 21~cm measurements \citep{Wilensky2019, Kolopanis2023}. The data might also be prone to uncertainties associated with the mitigation techniques, adding to the existing systematics.

This work embraces the first method where we reject any outlying observations. We developed a statistical framework that interrupts the data processing pipeline such that the output data products are thoroughly inspected before they proceed to the next step. It is designed such that any dysfunctional antennas, bad timestamps or frequency channels are discarded before the filtering process. After successful passing through the preliminary gateways, a set of filters are formulated from the derived metrics to identify outlying observations. However, the derived metrics may not be sufficiently robust to capture faint systematics \citep[e.g. ultra-faint RFI, ][]{Wilensky2023}.

The data processing pipeline is shown in Figure~\ref{fig:flowchart} whereby statistical metrics are derived and administered at the intermediate steps, with some of the main ones highlighted in red. The caveat of this strict approach is reduction in the number of observations that survives. Nevertheless we believe it is better to prevent systematics from escaping into the final measurements that would induce biases. We used observations from the MWA to implement the statistical framework. Details of the observational setup is presented in the next section.

\section{Observations}
\label{sec:observations}

The MWA is a radio telescope,  located at  \textit{Inyarrimanha Ilgari Bundara}, the 
Commonwealth Scientific and Industrial Research Organisation (CSIRO) Murchison Radio-astronomy Observatory, in the mid-west of Western Australia, about 300 kilometres inland from the coastal town of Geraldton. The location is considered pristine to study the evolution of our Universe for its low-level radio frequency interference \citep{Offringa2015}. The instrument serves as a precursor for the Low-Frequency Square Kilometre Array telescope, currently under construction on the same site.

\begin{figure}
	\centering
	\includegraphics[width=0.9\linewidth]{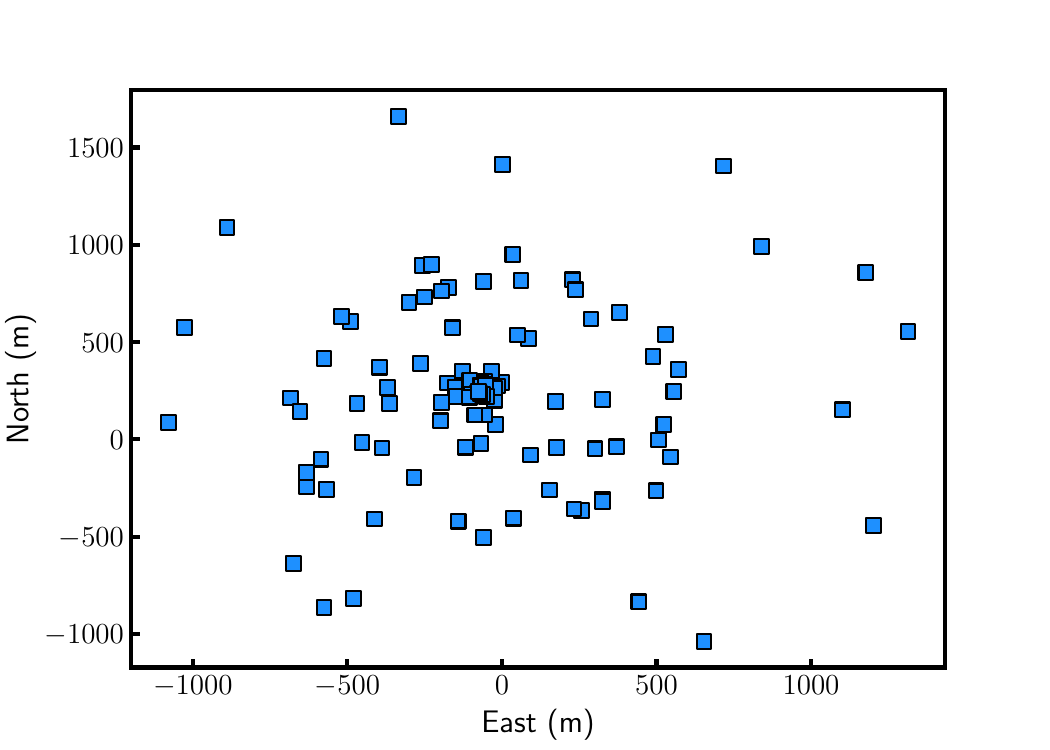}
	\caption{The telescope configuration during Phase I with 128 tiles pseudorandomly placed within a circumference of about 1.5$\sim$km.}
	\label{fig:ant_layout}
\end{figure}

The development of MWA is split into several phases. The instrument kicked off with 32 tiles in 2009 \citep[Phase 0,][]{Ord2010}, extending to 128 tiles in December 2012 \citep[Phase I; Figure~\ref{fig:ant_layout},][]{Tingay2013}. It was further upgraded to Phase II through the addition of 72 new tiles arranged in two compact hexagons along with 56 existing tiles pseudo-randomly spread (for detailed information refer to \citet{Wayth2018}). However, this work is restricted to observations from Phase I configuration. Each tile is made up of 4x4 dual polarized dipoles, optimized to operate between 80-300 MHz.

\begin{table}
	\begin{tabular}{c|c|c}
		\hline
		&  Altitude  &  Azimuth  \\
		\hline
		Pointing -3 & 69.2$^\circ$ & 90$^{\circ}$\\
		Pointing -2 & 76.3$^\circ$ & 90$^{\circ}$\\
			Pointing -1 & 83.2$^\circ$ & 90$^{\circ}$\\
			Pointing 0 & 90$^\circ$ & 0$^{\circ}$\\
		Pointing 1 & 83.2$^\circ$ & 270$^{\circ}$\\
		Pointing 2 & 76.3$^\circ$ & 270$^{\circ}$\\
		Pointing 3 & 69.2$^\circ$ & 270$^{\circ}$\\[0.5ex]
		\hline 
	\end{tabular}
	\caption{The altitudes and azimuths of the seven telescope beam pointings used in this analysis.  Pointing 0 is the zenith, -3 is three pointings before zenith, and 3 is three pointings after zenith.}
	\label{tab:pointing}
\end{table}

\begin{figure}[ht!]
	\centering
	\includegraphics[width=0.99\linewidth]{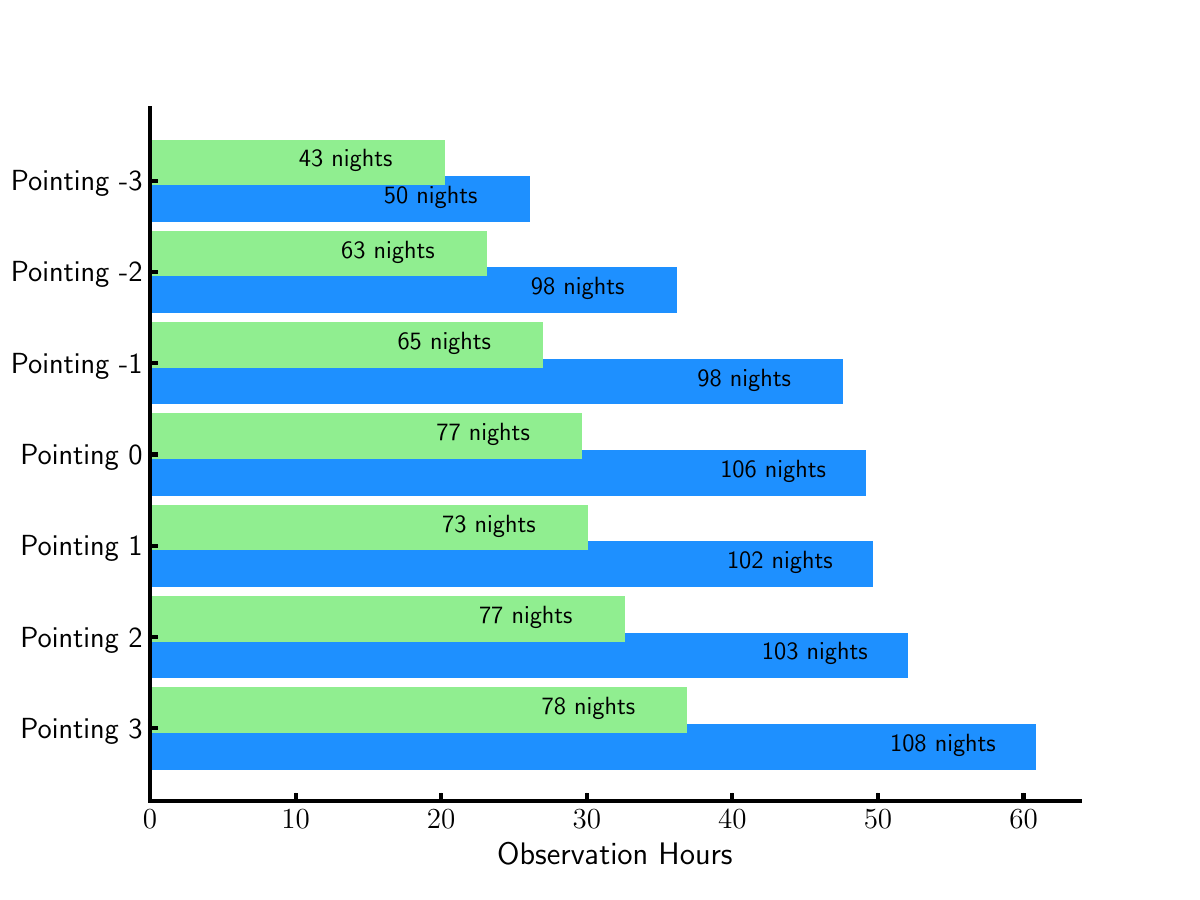}
	\caption{The bars in blue show the hours of observations extracted from MWA archival database for the individual pointing. The bars in green show the data that were selected based on the constraints described in Section~\ref{sub:data_quality}. Number of nights associated with the pointings are delineated on the bars.}
	\label{fig:obs_pointing}
\end{figure}

The telescope was steered at seven pointings listed in Table~\ref{tab:pointing}. In this paper, we will be targeting Phase I high band frequency observations between 167--197 MHz from EoR0 field centred at (RA=0~h, DEC=-27$^{\circ}$). The field consists of foregrounds contributed by the setting of the Galactic plane \citep{Barry2023} on the western horizon. The EoR experiment has observations spanning from 2013 to 2015 for the Phase I configuration. Each observation lasted for 2~minutes, totalling to about 322~hours (9655 in number). A breakdown of the data with respect to pointings are illustrated in Figure~\ref{fig:obs_pointing}.

%\section{Data Processing}\part{title}
\section{EoR Data Pipeline}
\label{sec:eor_pipeline}
\begin{figure*}
	\centering
	\includegraphics[width=0.85\linewidth]{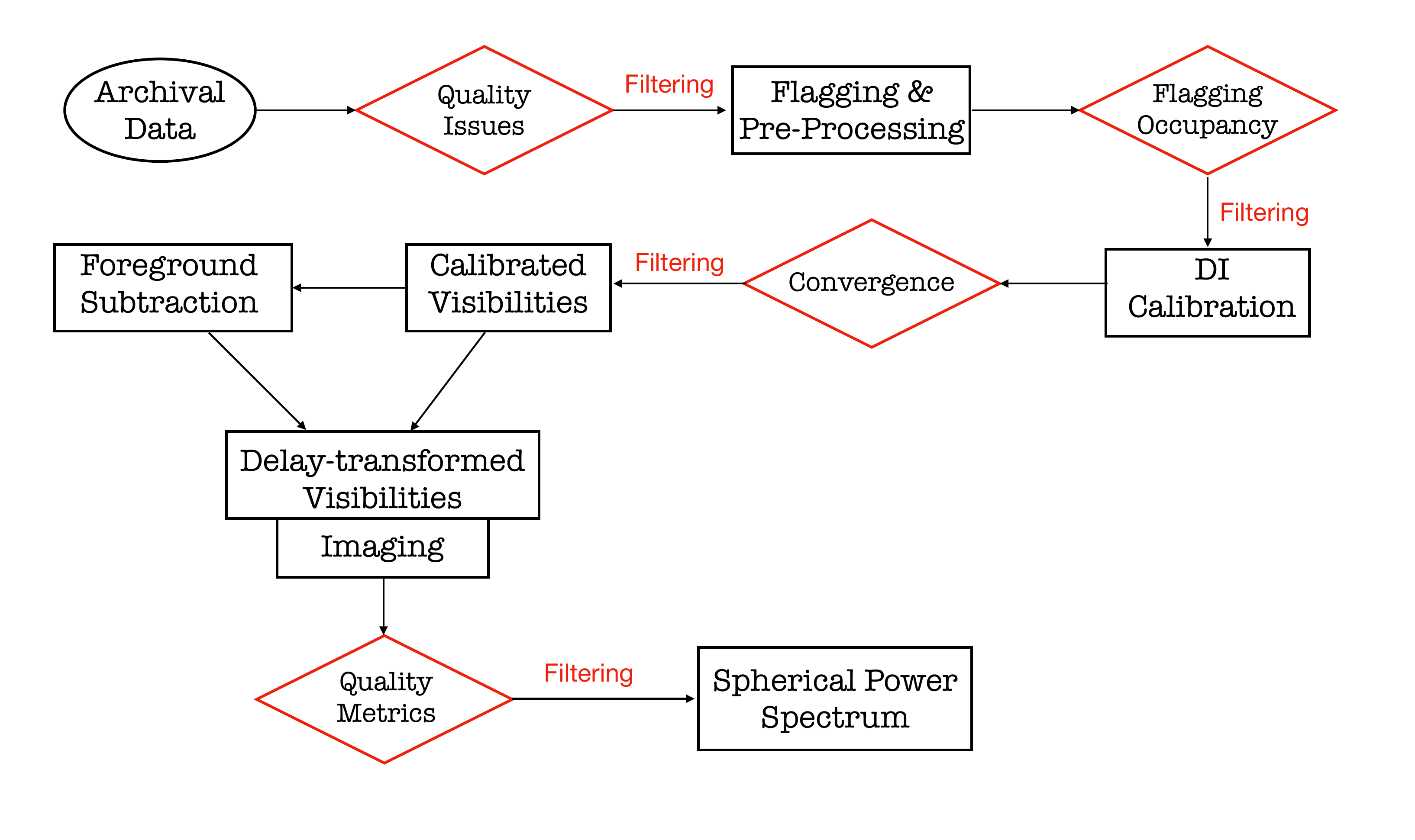}
	\caption{Data processing pipeline from extracting data to constructing a power spectra. The intermediate processes where inspection and filtering of the data products are carried out, are indicated in red. Each of the stages portrayed in black boxes are explained throughout the paper.}
	\label{fig:flowchart}
\end{figure*}

The data described in Section~\ref{sec:observations} were first downloaded from the MWA All-Sky Virtual Observatory (ASVO) database as raw files produced by the correlator. The downloaded data were then passed through the EoR pipeline. The boxes highlight the processes that include flagging, calibration, foreground subtraction, delay transform analysis, imaging, and power spectrum analysis. The data quality assessment procedures are denoted by the red rhombuses.
%to the processes highlighted in the blue boxes in Figure~\ref{fig:flowchart} through the EoR Pipeline
%A data selection was performed on the observations that were then passed to the flagging step followed by pre-processing, calibration, foreground subtraction and  imaging. 
%A flowchart is presented in Figure~\ref{fig:eor_pipeline} as guidance through the various stages involved in the pipeline.

\subsection{Flagging and Pre-processing}
\label{sec:data_reduction}
%An initial round of RFI excision is performed using \href{https://aoflagger.readthedocs.io/en/latest/}{AoLFlagger} \citep{Offringa2010, Offringa2012}. 
We applied \href{https://aoflagger.readthedocs.io/en/latest/}{AOFlagger} \citep{Offringa2010, Offringa2012} with the default MWA RFI strategy settings to the data. In addition to RFI identification and mitigation, AOFlagger flags known corrupted frequency channels namely the edges and centre of the coarse bands. Pre-processing was then conducted by \href{https://github.com/MWATelescope/Birli}{Birli} where the data was transformed from the correlator output format to a $UVFITS$ format. The frequency channels were averaged to $40$~kHz and the time intervals to $2$~s. 
The receiving signal chain undergoes a state change at the start of each observation, occurring $2$~seconds after the initiation of the  GPS time. Additionally, the ending timestamps may  potentially be affected by pointing, frequency or attenuation changes, rounded up to the next correlator dump time. Therefore, all timestamps $2$~seconds after the start and $1$~second before the end of each observation were flagged. The time flags vary across observations for several reasons: 1) only common timestamps of the coarse frequency channels were used; 2) some observations had late starting or early ending times; 3) averaging setting across time were different due to a mix of time resolutions in our observations, resulting in different weights being assigned. While averaging in frequency, the centre channel and $80$~kHz edge channels were flagged per $1.28$~MHz coarse band.

%leaving us with only one quarter of observations ($2161$; $72$~hours). This outcome is not surprising as \citet{Wilensky2019} found one third of the data used for the power spectrum in \citet{Beardsley2016} to be contaminated by DTV RFI. It is worth also noting that the difference in occupancies may be partly attributed to way both flaggers operate: AOFlagger identifies RFI on an antennas basis while SSINS performs its analysis on a per-baseline mode.

% Further flagging process is explained below.

We then focused on the autocorrelated visibilities, where the signal from one antenna is correlated with itself. Since we expect the bandpass gains to behave similarly across antennas, potential outliers can be identified from the autocorrelations before calibration. As the gains are stable across each observation ($2$ minutes interval), we averaged the autocorrelations in time. It is important to note that the autocorrelated visibilities were normalised by a reference antenna, which was taken to be last antenna in a non-flagged instrument configuration. Modified $z-scores$ were evaluated on the amplitudes of the averaged autocorrelations. Antennas with modified $z-$scores greater that $3.5$ were identified as outliers or dysfunctional antennas. The z-score analysis was iterated until no outliers were found. Subsequently, the dysfunctional antennas were flagged during calibration.

\subsection{Calibration}
\label{sec:calibration}
In a two-element radio interferometer, the correlation of the signal received at each element, termed as visibility, is measured. Assuming a flat sky, the relationship between the sky brightness distribution ${\bf S}$ and the visibility ${\bf V}$ across a baseline is given by

%${\bf v} = (v_{xx}, v_{xy}, v_{yx}, v_{yy})^T$ is given by
\begin{equation}
	\label{eq:measurement_equation}
 {\bf V}({\bf b}, \nu) = \int_{\Omega} {\bf A} ({\hat {\bf r}}, \nu)  {\bf S}({\hat {\bf r}}, \nu)  e^{-2\pi i \nu \frac{{\bf b}. {\hat {\bf s}}}{c}} ~d\Omega.
\end{equation}
Here, ${\bf b} = (u,v, w)$ represents the baseline projection, ${\hat {\bf r}}$ is the unit vector representing the direction cosines on the celestial sphere and $\nu$ is the observing frequency. %The Stokes vector ${\bf S}= (I+Q, U+iV, U-iV, I-Q)^T$ and 
The primary beam response of the antenna is denoted by the $2\times 2$ matrix:

\begin{equation}
{\bf A}=
\begin{pmatrix}
	A_{EW} & D_{EW}\\
	D_{NS} & A_{NS}
\end{pmatrix}
\end{equation}
where $A_x$ and $A_y$ represent the antenna response along the East-West (EW) and North-South (NS) directions respectively, and $D_{EW}$ and $D_{NS}$ are the terms that describe any instrumental leakage resulting from the signal from one polarisation escaping into the other \citep[see][]{Hamaker1996, Thompson2017}.
The signal received at the antenna gets corrupted along its propagation path by both direction independent and direction dependent antenna gains, thereby corrupting the visibilities. In this work, we solved for only the direction independent gains with \href{https://mwatelescope.github.io/mwa_hyperdrive/}{Hyperdrive} \citep{Jordan}, leaving the direction dependent gains for future.
%We solved for the direction dependant gains with \href{https://mwatelescope.github.io/mwa_hyperdrive/}{Hyperdrive} \citep{Jordan}. 
We used the MWA Long Baseline Epoch of Reionisation Survey (LoBES) catalogue as the foreground model, derived from the EoR0 field targeting EoR experiments \citep{Lynch2021}. Given that the model contains information only for the Stokes I parameter, the remaining three Stokes components in equation \ref{eq:measurement_equation} were assumed to be zero. We utilised the Full Embedded Element (FEE) primary beam model generated with Hyperbeam \citep{Sokolowski2017}.
\begin{figure}[h]
	\centering
	\includegraphics[width=\linewidth]{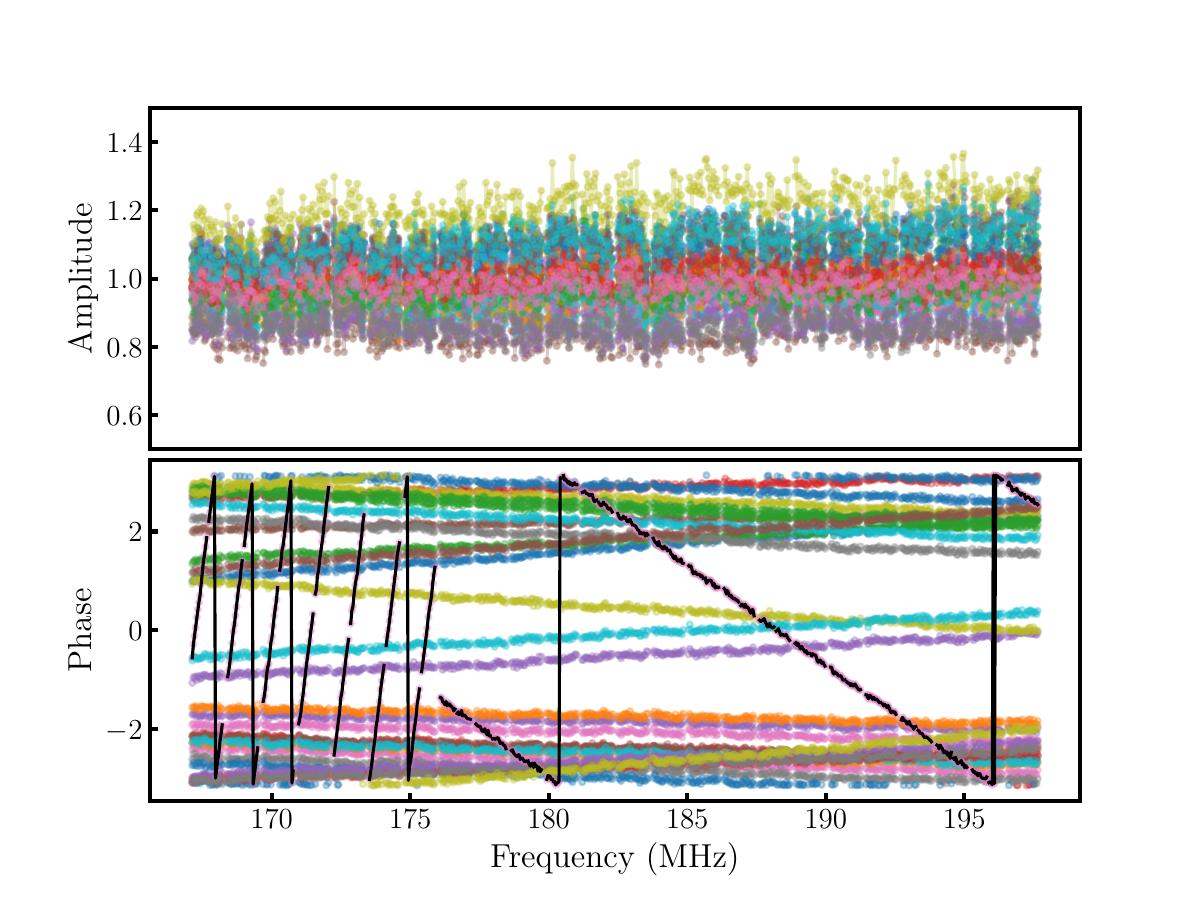}
	\caption{{\textit Top:} Amplitudes of the gain solutions obtained for an observation with the colours representing individual antennas. No misbehaving antenna are identified. {\textit Bottom:} Phases of the calibration solutions for the corresponding observation. The black points highlights the antenna with high fringing phases at low frequencies.}
	\label{fig:calibration_soln}
\end{figure}

\begin{figure*}[h!]
	\centering
	\includegraphics[width=\linewidth]{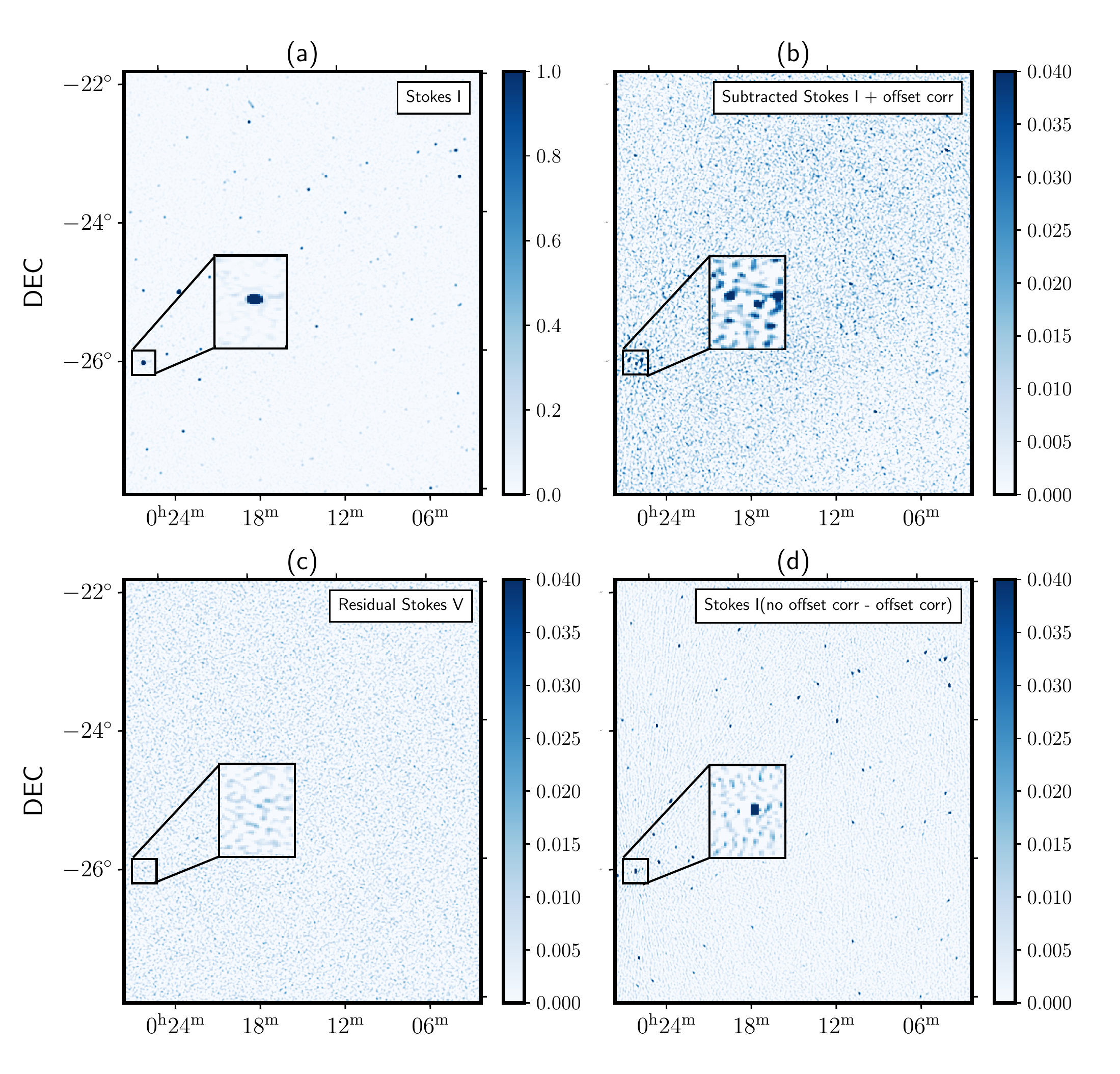}
	\caption{\textit{Left}: Deconvolved images generated through the process described in Section~\ref{sec:imaging} for a zenith pointing observation with (a) representing Stokes $I$ formed from unsubtracted visibilities, (b) and (c) representing Stokes $I$ and $V$ from subtracted and ionospheric phase corrected visibilities respectively and (d) difference between source subtracted visibilities before and after phase offset correction. The units of the colorbar is Jy beam$^{-1}$. The inset shows a zoom version of the region surrounding the brightest source in the field of view $PKS0026-23$.}
		%Plots (a) and (c) present the images along $xx$ and $yy$ polarisations respectively while (e) shows the  Stokes $V$ image. The RMS values for (a), (c) and (e) are 26.45, 28.16 and 0.30~Jy~beam$^{-1}$ respectively. \textit{Right}: Same as the left panel but after standard subtraction (before phase offset correction) with (b) and (d) showing the $xx$ and $yy$ polarisations respectively and (f) Stokes $V$. The units of the colorbar is Jy beam$^{-1}$}
	\label{fig:stokes_fitsfile}
\end{figure*} 

The calibration process involves applying least square minimization, iteratively solving for per-antenna complex gains for each frequency and time, enabling the capture of spectral structures. With the number of iterations set to 50 and the convergence threshold at $10^{-6}$, the solutions encapsulated most spectral features. Ideally, a convergence closer to zero is desirable, but this would require increasing the number of iterations, thus increasing computational requirements. As this work focuses on the diagnosis side of the data processing pipeline, the convergence aspect falls beyond the scope of this paper. For each antenna, frequency channels that did not reach the specified threshold were flagged within the calibration process, possibly due to RFI or systematics.

We then investigated the calibration solutions for anomalies. The gain solutions were normalised with respect to the last unflagged antenna. We began by evaluating the RMS of the gain amplitudes and antennas and a three sigma thresholding was applied for identifying misbehaving antennas. However, as depicted in Figure~\ref{fig:calibration_soln}, we were not able to spot poor behaviour such as fast fringing of the phases at low or high frequencies from the amplitudes. These behaviours were hence, manually identified and removed.

The solutions were then applied to form calibrated visibilities that were fed into foreground mitigation step.

%%%%%%%%%%%%%%%%%%%%%%%%%%%%%%%%%%%%%%%
\subsection{Foreground Subtraction}
\label{sec:foreground_subtraction}

In this work, we employed a foreground subtraction approach, wherein we modelled and subtracted foreground visibilities for 4000 sources within the field of view. These model visibilities were constructed following the methodology outlined in Section~\ref{sec:calibration}, utilizing the LoBES catalogue and FEE primary beam. This method effectively decreased foreground contamination at low $k$ modes by approximately an order of magnitude in the EoR window.
However, our analysis revealed shortcomings in the results obtained through standard subtraction techniques, manifesting as either under-subtraction or over-subtraction. This behavior can primarily be attributed to direction-dependent effects caused by the ionosphere, leading to positional phase shifts. Such phenomena have been previously investigated in MWA observations and addressed through the `peeling' technique, described in \citep{Mitchell2008}.

Following suit, we incorporated this peeling approach into our analysis, correcting phase offsets for the 1000 brightest sources in the field of view. The selection of sources for peeling was determined based on the minimal requirement of approximately $50-60$ sources for an image size of $\sim 30^{\circ}$, as evaluated in \citep{Mitchell2008}. However, the computational resources imposed a cap on the number of sources that could be included.
We evaluated the efficacy of this peeling technique through imaging improvements, which will be discussed in the following subsection.

\subsection{Imaging}
\label{sec:imaging}

The next step in our data processing pipeline is generating images to reinforce the quality assurance through visual inspection and add to the statistical metrics. Images with angular resolutions of $40$~arcseconds were formed by Fourier transforming the visibilities along the East-West and North South directions (EW and NS polarisations) using WSClean \citep{Offringa2014}. All the sub bands were combined together using the multifrequency synthesis algorithm. Briggs weighting with robustness of $-1$ was used such that we emphasised more on the resolution and reduction of the sidelobes but at the same time increasing the signal to noise ratio for quality assurance \citep{Briggs1995}. % The dirty images were deconvolved through multiclean strategy whereby a mask of the cleaning region was defined.
 Cotton–Schwab algorithm was employed and the images were deconvolved down to a threshold of $1$~Jy, chosen to reduce computational cost as deeper cleaning was not required for diagnostic purpose. Example of a pseudo-Stokes $I$ image is shown in plot (a) of  Figure~\ref{fig:stokes_fitsfile}. Stokes $V$ images were also created in the same fashion.

\begin{figure}
	\centering
	\includegraphics[width=\linewidth]{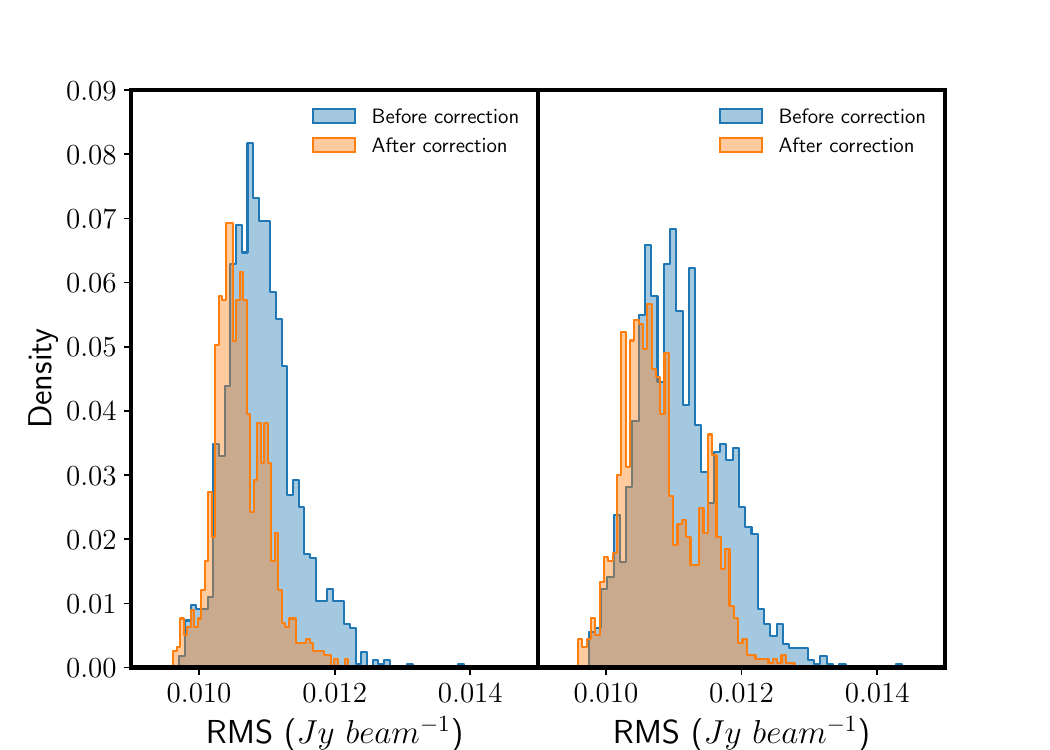}
	\caption{RMS distribution across all observations before and after phase offset correction for 
		EW (left) and NS (right) polarisations.}
	\label{fig:hist_ionosubvssub}
\end{figure}

Panel (b) of  Figure~\ref{fig:stokes_fitsfile} were formed from the subtracted visibilities that were corrected for ionospheric phase offset. The overall RMS in Stokes $I$ drops from $1.66$ Jy~beam$^{-1}$ to $0.5$ Jy~beam$^{-1}$. We found that subtraction across EW polarisation performed better with a decrease in RMS by a factor of 60 while for NS, the factor is 23 because of the galactic plane aliasing being more prominent along this direction. Stokes $V$ plotted in panel (c) showed marginal difference after subtraction. The improvement made to the subtraction after accounting for phase offsets is illustrated in the difference image in (d). It is observed that apart from a better subtraction of the brightest source in the field of view  $PKS0026-23$ resulting into reduced sidelobe intensities, there are other visible sources that were under subtracted. We also found a flux difference of 900~mJy in $PKS0026-23$. The overall RMS for this observation drops by an order of magnitude for EW polarisation while marginal difference is found in NS and no difference in Stokes $V$. The RMS across all the pixels for each observation  is plotted in Figure~\ref{fig:hist_ionosubvssub}. The mean is seen to be shifted slightly to the left after correction and the standard deviation is more constrained for both polarisations. These pointers indicate a significant refinement in the foreground removal.

After foreground subtraction, observations were fed into the power spectrum machinery.

	\section{Power Spectrum Analysis}
	\label{sec:power_spectrum}
	The power spectrum defines the power of the signal as a function of $k$ modes. In $k$ space, $(u, v)$ represents the Fourier modes of the measured 
	visibility points, $(l, m)$ are the angular modes $k_{\perp}$ and $k_{||}$ are modes paralell to the line-of -sight mapped from the spectral channels. The power spectrum can then be given by
		\begin{align}
		P(k)  &= P(\sqrt{k_{\perp}^2 + k_{\parallel}^2})  \nonumber \\
		         & = \frac{1}{\Omega}\langle \tilde{{\bf V}}(k) \tilde{{\bf V}}^*(k)\rangle
	   \end{align} 
	   where $\Omega$ is the observing volume. The visibilities in Equation~\ref{eq:measurement_equation} are gridded onto a $uv$ grid and Fourier transformed along the frequency axis. The one-dimensional power spectra can hence be defined as the integrated power over $k$ space:
	   \begin{equation}
	   	\Delta^2 = \frac{k^3}{2\pi^2} P(k).
	   \end{equation}

%	The key difference amongst the implementations of the power spectra is the approach to which the visibilities are treated. HERA relies on the drift scan ability of the instrument and use a delay-transform approach while MWA has been approached from various angles, namely the Fast Holographic Deconvolution \citep[FHD/eppsilon;][]{Sullivan2012, Barry2019}, the simpleDS \citep{Kolopanis2023} and Cosmological HI Power Spectrum estimator \citep[CHIPS;][]{Trott2020}. The simpleDS pipeline was implemented fundamentally for redundant configurations, targetting Phase II MWA data. FHD/eppsilon combines data in the image space while CHIPS directly grids the visibilities into a $(u, v, \nu)$. 
 In our work, we utilised the Cosmological HI Power Spectrum estimator (CHIPS) pipeline for generating power spectra from MWA observations \citep{trott2016}. While we did not perform a direct comparison or validation of our results with other approaches, we believe it is valuable to explain our choice.
	
	Among the active pipelines for MWA data power spectrum generation, including Fast Holographic Deconvolution \citep[FHD/eppsilon;][]{Sullivan2012, Barry2019} and simpleDS \citep{Kolopanis2023}, we opted for CHIPS due to several key considerations. Firstly, we ruled out the simpleDS pipeline because it is primarily designed for redundant configurations, targeting Phase II MWA data, which did not align with our observational setup.
	CHIPS constructs power spectra from a discrete $uv$-plane, whereas FHD/eppsilon adopts an image-based procedure. By utilizing the uvw plane, CHIPS effectively circumvents aliasing issues that FHD/eppsilon encounters. Moreover, CHIPS applies an inverse variance weighting, to account for the frequency-dependent weights in an optimal way.
Previous studies \citep{Barry2019} have demonstrated that the results obtained from both CHIPS and FHD/eppsilon pipelines exhibit consistency and follow a general trend. \citet{line24} also demonstrates that the Hyperdrive-CHIPS pipeline does not suffer from signal loss. This further supports our decision to utilise CHIPS for our analysis, ensuring robust and reliable power spectrum estimation from MWA observations.

%	{\bf There are currently few active pipelines that uses MWA observations to generate power spectra, namely, the Fast Holographic Deconvolution \citep[FHD/eppsilon;][]{Sullivan2012, Barry2019}, the simpleDS \citep{Kolopanis2023} and Cosmological HI Power Spectrum estimator \citep[CHIPS;][]{Trott2020}. We used CHIPS for this work. While we do not perform any comparison or validation of our results with other approaches to generate power spectrum, we find it useful to explain our choice.  We ruled out the simpleDS pipeline as it was implemented fundamentally for redundant configurations, targetting Phase II MWA data. CHIPS constructs its power spectra from discrete $w$-projected plane, while FHD/eppsilon adopts an image-based procedure. By using the $uvw$ plane, CHIPS escapes the aliasing issues that FHD/eppsilon has to deal with \citep{Barry2019, Barry2024}. Additionally, CHIPS addresses systematics through inverse covariance weighting \citep{Kay1993}, allowing to probe low $k$ modes, regions prone to aliasing harmonics. \citep{Barry2019} demonstrated that the results from both pipeline are consistent and followed a general trend.}
%		
	%The comparison between FHD/eppsilon and CHIPS were conducted in \citet{Liu2019}.
	 %This work adopts the CHIPS methodology to estimate the power spectra.
	%%%%%%%%%%%%%%%%%%%%%%%%%%%%%%%%%%%

	\section{Data quality assurance}
	\label{sec:data_assurance}
	A data quality assessment was conducted at the stages outlined in Figure~\ref{fig:flowchart}, enabling us to identify and filter visibilities exhibiting anomalous behaviour and patterns. These anomalies were identified from various data products, as discussed in the following subsections. The cutoff thresholds and number of successful observations that passed through the various stages are presented in Table~\ref{tab:metrics}.	
	
		\subsection{Data Quality Issues}
	\label{sub:data_quality}
	The archival data from Section~\ref{sec:observations} were triaged before pre-processing as described in Section~\ref{sec:data_reduction}, to ensure that no data quality uncertainties were associated with them. Elements considered in this process included:
	\begin{itemize}
		\item Errors in beamformer communication on individual tiles.
		\item Discrepancies in attenuation settings of the receiver.
		\item Recorded events from the Monitor and Control System.
		\item Presence of two or more disabled dipoles along the same polarisation, indicating a flagged tile.
	\end{itemize}
	
\begin{figure}
		\centering
		\includegraphics[width=0.99\linewidth]{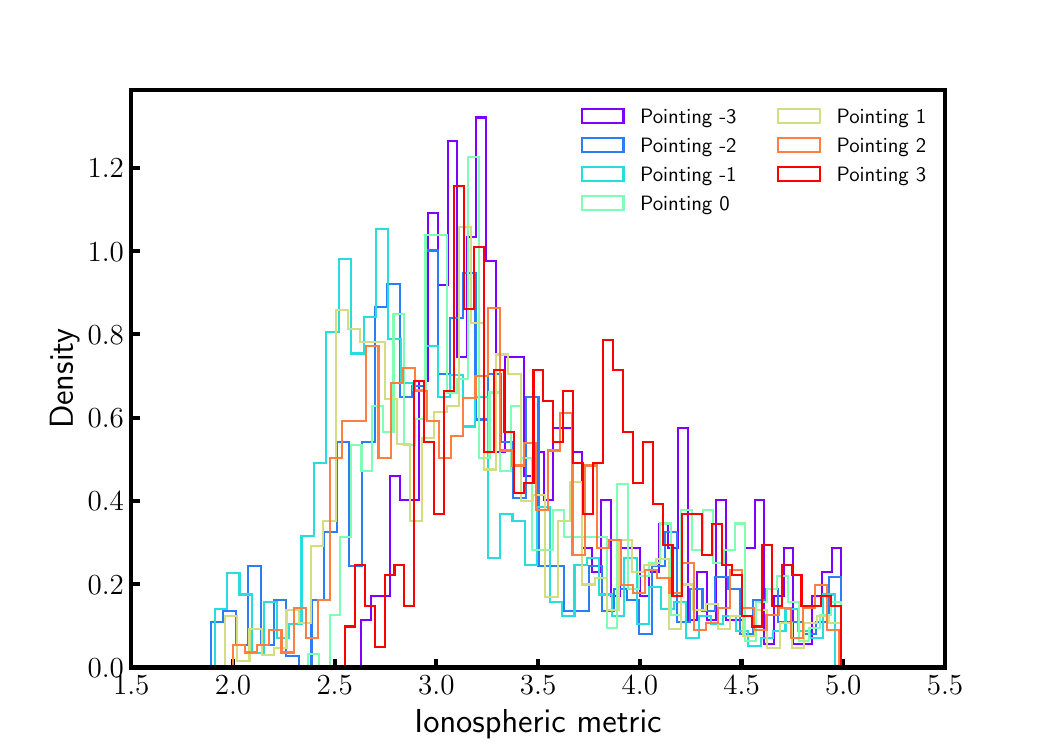}
		\caption{The ionospheric distribution after applying the threshold cut-off of $5$. The colors represent the different pointings.}
		%Despite the variation of skewness and kurtosis of the distribution across pointings, the standard deviation remains fairly constant.}
	\label{fig:ionoqa_hist}
	\end{figure}

Additionally, observations with high levels of ionospheric activity, capable of distorting our measurements, were identified and discarded. This was achieved using the ionospheric metric developed by \cite{Jordan2017}, which incorporates the median source offset and source offset anisotropy derived from the measured versus expected source positions of 1000 point sources in the field of view. Observations yielding an ionospheric metric greater than the cut-off threshold estimated in \cite{Trott2020} were excluded, resulting in 4943 observations (equivalent to 165 hours), as depicted by the green area in Figure~\ref{fig:obs_pointing}. The ionospheric distributions are illustrated in Figure~\ref{fig:ionoqa_hist}, with mean values ranging from 3.8 to 4.5 arcminutes across pointings. This metric serves as a proxy for ionospheric activity in an observation, with lower values indicating less ionospheric activity. The Kolmogorov-Smirnov test indicates that these distributions are not normally distributed.

	\subsection{Flagging Occupancy}
		\begin{figure*}
		\centering
		\includegraphics[width=1\linewidth]{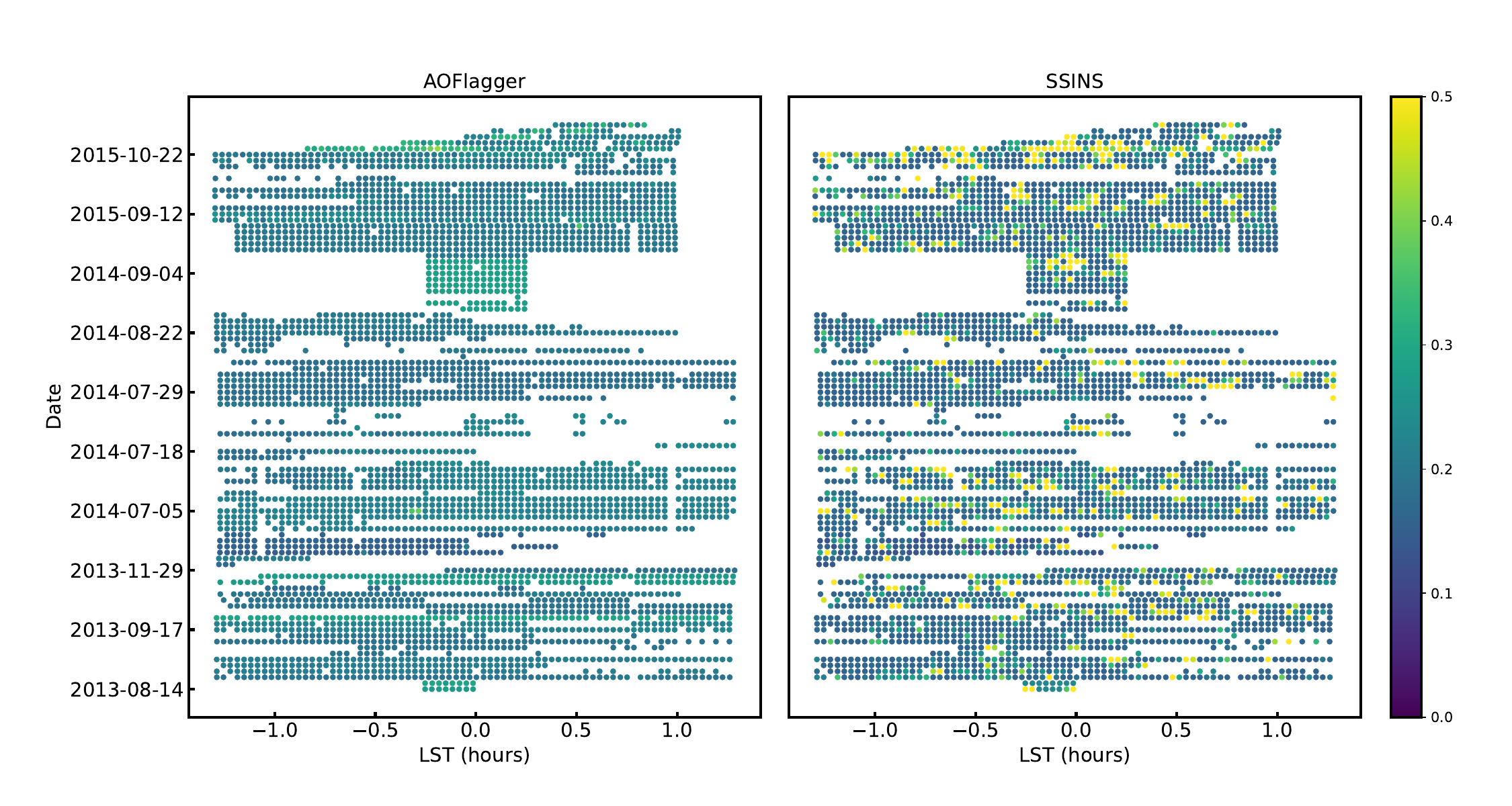}
		\caption{{\textit Left:} The total occupancy obtained from the pre-flags and AOFlagger, described in Section~\ref{sec:data_reduction}. The flags are displayed only for a portion of the observations to avoid crowding and across the Local Sidereal Time (LST). The negative LSTs should be read as the corresponding LST subtracted from 24~hours. {\textit Right:} Same as the left panel with the total occupancy evaluated from SSINS flagger \citep{Wilensky2019}.}
		\label{fig:aol_occupancy}
	\end{figure*}

	The flagging occupancy was calculated based on flags generated from data quality issues outlined in Section 6.1, as well as flags obtained through the application of AOFlagger and autocorrelation analysis on the observations (refer to Section ~\ref{sec:data_reduction}). The left panel of Figure Figure~\ref{fig:aol_occupancy} illustrates the flagging occupancy for a set of MWA observations. Observations with a flagging occupancy greater than 25$\%$ were discarded.
	
	However, studies by \citet{Wilensky2019} and \citet{Barry2019} revealed that AOFlagger may overlook faint systematics, particularly faint RFI residing below thermal noise. Hence, we further evaluated the flagging occupancy on the flagged visibilities to assess the quality of our data. Notably, flags produced by SSINS were used solely for the filtering process.
	
	The SSINS flagger provides occupancies for four classes of RFI: faint broadband streak, narrow broadband interference, DTV Signal, and total occupancy. Faint broadband refers to systematics of unknown origins occupying a wide band, while narrow interference relates to systematics at a specific frequency or a narrow band. DTV signal represents the full propagation of the DTV interference across all baselines, partly identified by AOFlagger, and total occupancy evaluates the underlying faint systematics identified by the algorithm over the entire observation, as illustrated in the right panel of Figure~\ref{fig:aol_occupancy}.
	
\begin{figure}
		\centering
		\includegraphics[width=1\linewidth]{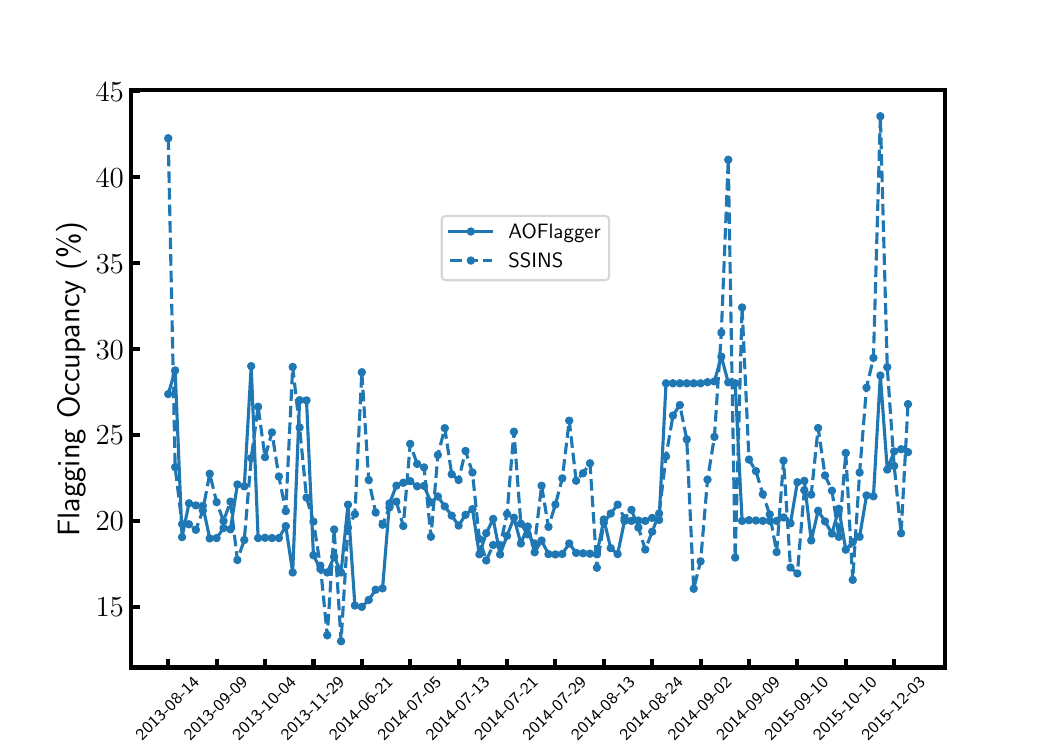}
		\caption{Flagging occupancy evaluated by AOFlagger (solid) and SSINS (dotted) averaged over observation for individual nights.}
		\label{fig:aol_occupancy_total}
	\end{figure}

	Observations exhibiting any broadband, narrow interferences, or DTV signal were discarded. The averaged flagging occupancy for each night was evaluated for both flaggers, shown in Figure~\ref{fig:aol_occupancy_total}. It serves as another comparison between the occupancies flagged by AOFlagger and SSINS, highlighting faint systematics overlooked by AOFlagger, but identified by SSINS. These faint systematics may originate from faint RFI, sources at the horizon attenuated enough by the primary beam to evade detection by AOFlagger, or other unknown RFI origins. Observations with SSINS occupancy exceeding 25$\%$ were rejected, resulting in only one quarter of observations remaining (2161; 72 hours). This outcome aligns with the findings of  \citet{Wilensky2019}, who reported that one third of the data used for the power spectrum in \citet{Beardsley2016} was contaminated by DTV RFI. It is also noteworthy that the difference in occupancies may partly be attributed to the way both flaggers operate: AOFlagger identifies RFI on an antenna basis, while SSINS performs its analysis on a per-baseline mode.

	\subsection{Calibration Solutions}
	The quality of each observation was further assessed concerning the calibration solutions derived in Section~\ref{sec:calibration}. The least-squares minimization algorithm yielded convergence values, indicating the degree to which the solutions approached zero. We employed the variance of these convergence values across frequencies as a deviation metric. Observations with deviations surpassing the square root of the specified stopping threshold were flagged as outliers and treated accordingly.

	%did not yield reliable solutions (e.g., noise-dominated solutions, unusual spectral structures across amplitudes, and phase irregularities).
	
%
%	Convergence values were further used as a metric to determine observation reliability. The variance of the convergence values, measuring the deviation of convergence across a particular frequency channel from the mean, was weighed against the convergence threshold. Upon experimentation, observations with a convergence variance exceeding half the specified convergence threshold did not yield reliable solutions (e.g., noise-dominated solutions, unusual spectral structures across amplitudes, and phase irregularities). Therefore, this criterion was applied for discarding observations. 
%	Further quality assessment on the observations were performed before they were passed onto the power spectrum analysis. In addition to the filters included throughout the pipeline, we enforced four data quality metrics:

\subsection{Delay-transformed power spectrum}
\begin{figure*}
	\centering

	\includegraphics[width=\linewidth]{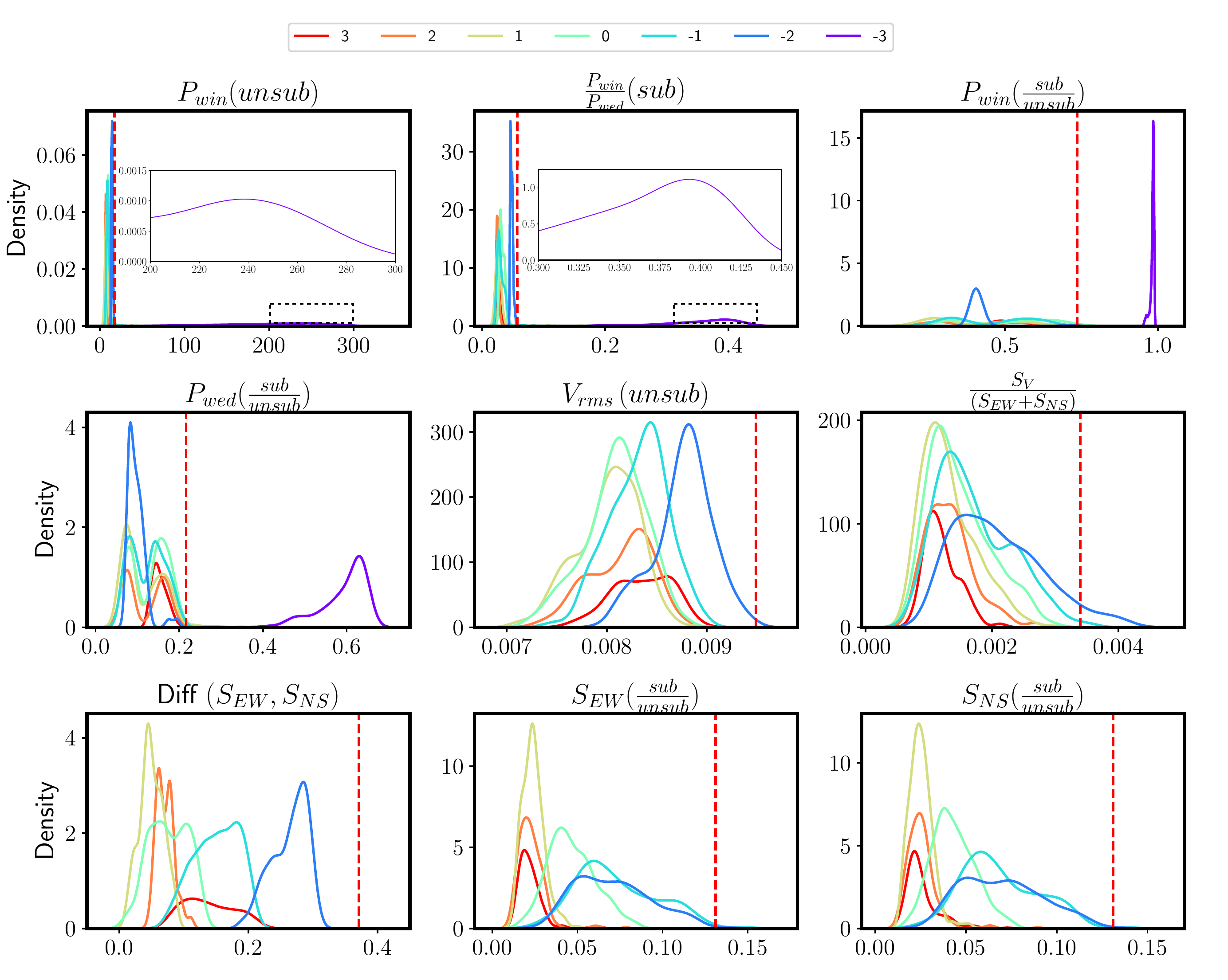}
	\caption{\textit{Top}: RMS values of Stokes $V$ in Jy~beam$^{-1}$ of the chosen observations plotted as a function of LST (adopting same LST convention as Figure~\ref{fig:aol_occupancy}). The colours indicate pointing. \textit{Bottom} Ratio of estimated $PKS0026-23$ flux density from Stokes $V$ to sum of EW and NS images as a function of LST.}
	\label{fig:filter_cuts}
\end{figure*}

The delay transformed power spectrum estimator \citep{Beardsley2016} were used to derive:
\begin{itemize}
	\item Power Spectra Wedge Power $P_{\textrm{wed}}$: The power  in Jy$^2$ confined within the area underneath the boundary set by the chromaticity of the instrument for baselines $< 100\lambda$ and averaged by the number of contributing cells. 
	\item Power Spectra Window Power $P_{\textrm{win}}$: The power in Jy$^2$ in the EoR window up till the first coarse channel, corresponding to $k_{||} < 0.4~h$~Mpc$^{-1}$. Baselines $< 100 \lambda$ were included in the averaging and the power was normalized by the number of contributing cells \citep{Beardsley2016}.
\end{itemize}
We then constructed four data quality metrics with the above quantities to assess our observations and these are:
\begin{enumerate}
	\item $P_{\textrm{win}} \, (\textrm{unsub})$: Window power $P_{\textrm{win}}$ evaluated from the delay-transformed visibilities before foreground subtraction. %It provides us with a quantitative value of the foreground leakage beyond the wedge.
	\item $\frac{P_{\textrm{win}}}{P_{\textrm{wed}}} \,(\textrm{unsub})$:  Ratio of window power to wedge evaluated from the delay-transformed visibilities before foreground subtraction.% It measures the efficacy of our foreground removal process by quantifying the maximum power being subtracted.
	\item $P_{\textrm{win}} (\frac{\textrm{sub}}{\textrm{unsub}})$: Ratio of window power evaluated from the delay-transformed visibilities after foreground subtraction to before subtraction. %It measures the amount of power subtracted in the window region.
	\item $P_{\textrm{wed}}(\frac{\textrm{sub}}{\textrm{unsub}})$: Ratio of wedge power evaluated from the delay-transformed visibilities after foreground subtraction. % Same as  $P_{\textrm{win}} (\frac{\textrm{sub}}{\textrm{unsub}})$, but constrained to the wedge region.
\end{enumerate}
Given the widely spread distribution of the above mentioned quantities, illustrated in Figure~\ref{fig:filter_cuts}, particularly the distribution of pointing $-3$, we adopted the interquartile range as it is known for being resilient to extreme values. The derived cut-off thresholds, and the corresponding  number of successful observations are presented in Table~\ref{tab:metrics}. Observations that portray sufficient leakage into the window, with a maximum power value of 17~Jy$^2$ were thrown. 
If the ratio of the maximum power in the window to the wedge were greater than $5.7\%$, the observation were treated as an anomaly. 
The power removed by our subtraction methodology were quantified using the ratio of maximum power measured after foreground subtraction in both wedge and window regions to the power measured before subtraction. If the ratio resulted in values greater than $21\%$ and $73\%$ for the wedge and window respectively, the observation was excluded. As a result, all observations from pointing $-3$ were filtered out. 
%In this case, the distributions shown in Figure~\ref{fig:psfilter_cuts} are widely spread, therefore the $3$-sigma rule does not provide reliable results. We adopted the interquartile rule whereby observations below $1.5 \times$ the first {\bf quartile} and above $1.5 \times$ the third {\bf quartile} were rejected.
\begin{table}
	\caption{List of metrics used as diagnosis for filtering observations.}
	\label{tab:metrics}
	\begin{tabular}{c|c|c}
		Metric & Threshold cutoff & No. of observations \\
		\hline \hline
		Data quality issues &  0 & 218\\
		Ionospheric metric &   5 & 4494\\
		AoFlagger Occupancy &  25$\%$ & 440 \\
		SSINS Occupancy & 25$\%$ & 2342\\
		$P_{\textrm{win}} \, (\textrm{unsub})$ & 17~Jy$^2$ & 337\\
		$\frac{P_{\textrm{win}}}{P_{\textrm{wed}}} \,(\textrm{unsub}) $& 5.7\% & 24 \\
		$P_{\textrm{win}} (\frac{\textrm{sub}}{\textrm{unsub}})$ & 73\% & 53\\
		$P_{\textrm{wed}}(\frac{\textrm{sub}}{\textrm{unsub}})$ & 21\% & 0\\
		$V_{\textrm{rms}} (\textrm{unsub})$  & 9.5~mJy & 1 \\
		$\frac{S_{V}}{(S_{\textrm{EW}} + S_{\textrm{NS}})}$:& 0.3\%& 7	\\
		$(S_{\textrm{EW}} , S_{\textrm{NS}})$ & 37$\%$ & 0 \\
		$S_{\textrm{EW}}(\frac{\textrm{sub}}{\textrm{unsub}})$ & 13\%& 5 \\
		$S_{NS}(\frac{\textrm{sub}}{\textrm{unsub}})$ & 13$\%$& 0 \\
		\hline
	\end{tabular}
\end{table}

\subsection{Images}
The images generated from the imaging process described in Section~\ref{sec:imaging} were used to determine the following:
	\begin{itemize}
		\item RMS of Stokes $V$ image: The root mean square over a small area ($100$ by $100$ pixels area) sitting at one corner of the image, away from the concentration of source emissions.
		%over all the pixels in Stokes $V$ image generated using the parameters mentioned in Section~\ref{sec:imaging}. 
		\item $PKS0026-23$ flux density $S$: $PKS0026-23$ is the brightest source in the field of view with a flux density of  17.47~Jy at 150~MHz as reported in the GLEAM survey \citep{Wayth2015, HurleyWalker2017}. Here, a naive extraction of the flux density was done. It was evaluated as the integration over the area centred at the source within a radius spanning two synthesized beams. The radius was chosen to account for any remaining phase offset. The extraction was done separately on both polarisations.
	\end{itemize}

Using the aforementioned quantities, five data quality assessment metrics were constructed:
%As discussed in the paper, filters applied were based on the cut-off thresholds imposed on the different metrics. The quantities formed from the images are:
	\begin{enumerate}
		\item $V_{\textrm{rms}} (\textrm{unsub})$ : RMS across a selected pixel box in Stokes $V$ image. 
			\item $\frac{S_{V}}{(S_{\textrm{EW}} + S_{\textrm{NS}})}$:  Ratio of $PKS0023-026$ flux density $S$ extracted from Stokes $V$ to sum of flux density extracted from EW and NS polarisations.
			\item $(S_{\textrm{EW}} , S_{\textrm{NS}})$: Difference between $PKS0023-026$ flux densities across EW and NS polarisations.
			\item $S_{\textrm{EW}}(\frac{\textrm{sub}}{\textrm{unsub}})$: Ratio of $PKS0023-026$ flux density from subtracted image to unsubtracted image along EW polarisation.
			\item $S_{NS}(\frac{\textrm{sub}}{\textrm{unsub}})$:  Ratio of $PKS0023-026$ flux density from subtracted image to unsubtracted image along NS polarisation.
	\end{enumerate}
	
\begin{figure}
		\centering
		\includegraphics[width=\linewidth]{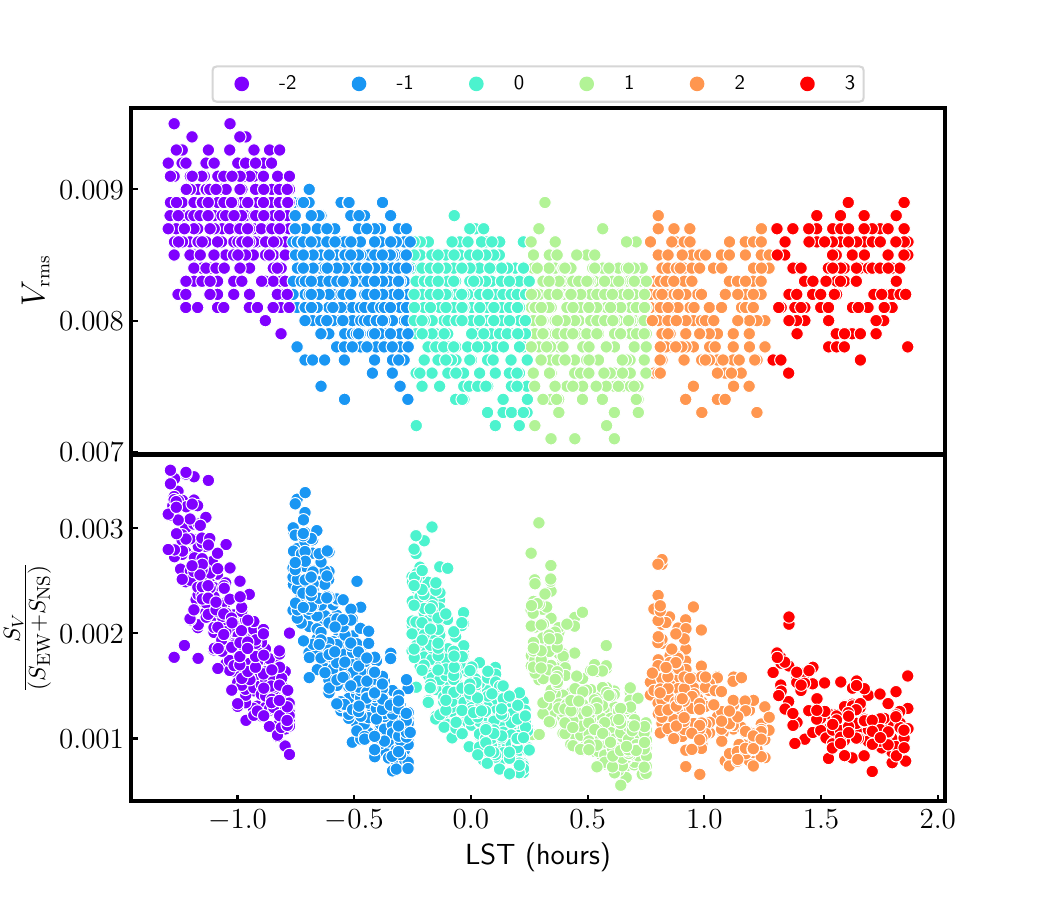}
		\caption{\textit{Top}: RMS values of Stokes $V$ in Jy~beam$^{-1}$ of the chosen observations plotted as a function of LST (adopting same LST convention as Figure~\ref{fig:aol_occupancy}). The colours indicate pointing. \textit{Bottom} Ratio of estimated $PKS0026-23$ flux density from Stokes $V$ to sum of EW and NS images as a function of LST.}
		\label{fig:image_qa}
	\end{figure}

The cutoff thresholds for the derived quantities were evaluated using the $3\sigma$-rule. The results are provided in Table~\ref{tab:metrics}. Since the Murchison Widefield Array (MWA) consists of linearly polarized dipoles, we anticipate minimal circularly polarized visibilities, as there are no bright Stokes V sources within our primary beam. The distribution of the pixels in Stokes $V$ should, in principle, be noise-like; therefore, the mean is expected to be around zero, with no skewness. Observations not adhering to this criterion and bearing an RMS value greater than $9.5$ mJy are filtered out. The top panel of Figure~\ref{fig:image_qa} presents the RMS values in Stokes $V$ for observations that passed this criterion as a function of local sidereal time. We observed that this filter excludes all observations from pointing $-3$. It is evident that the East-West negative pointings exhibit higher RMS values than the positive ones. Pointings $-2$ and $-1$ have the Galactic Centre in the second sidelobe of the primary lobe, and the emission could be leaking into Stokes $V$. The increase in RMS towards the positive pointings may be attributed to the influence of Fornax A, as it moves into the first sidelobe of the primary beam.

%	Here, the cutoff thresholds for the derived quantities were evaluated using the $3\sigma$-rule. The results are given in Table~\ref{tab:metrics}.	Given MWA consists of linear polarized dipoles, we expect to see very little circularly polarized visibilities as there are no bright Stokes V source within our primary beam.
%	The distribution of the pixels in Stokes $V$ should in principle be noise-like, therefore the mean is expected to be around zero, with no skewness. Observations not abiding by the the aforementioned criterion and bearing RMS value greater than $9.5$~mJy are filtered out. Top panel of Figure~\ref{fig:image_qa} presents the RMS values in Stokes $V$ for observations that passed this gateway as a function of local sidereal time. We noticed that this filter excludes all observations from pointing $-3$. It can be seen that the East-West negative pointings have higher RMS values than the positive ones. Pointings $-2$ and $-1$ have the Galactic Centre in the second sidelobe of the primary lobe and the emission could be leaking into Stokes $V$. The increase in RMS towards the positive pointing may be due to the impact of Fornax~A, as moves into the first sidelobe of the primary beam.
%	
	
	The flux density ratio of the source $PKS0023-026$ in Stokes $V$ to the sum of EW and NS polarisations (equivalent to pseudo-Stokes $I$) was calculated. This quantity informs us about the percentage of instrumental leakage from $(EW+NS) \rightarrow V$ that could also occur in the reverse direction, $V \rightarrow (EW+NS)$. Observations with an estimated leakage greater than $0.3\%$ were discarded. Fractional ratios for the successful observations are shown in the bottom panel of Figure~\ref{fig:image_qa}, following a similar trend as the RMS, except for pointings 2 and 3. The flux densities of $PKS0026-23$ from Stokes V images conform to the RMS trend, except for pointings 2 and 3. This behavior might be attributed to the position of $PKS0026-23$, such that Fornax~A has a negligible contribution.
	The ratio of the flux density after to before foreground removal was also scanned, such that observations with values below $13\%$ were allowed to proceed.
	
	The aforementioned metrics aimed to mitigate observations dominated by RFI, ionosphere, or contamination from nearby bright emissions. They also identified observations for which calibration and foreground subtraction did not perform well. This under-performance may be attributed to unidentified or unclassified systematics.

	\section{Results}
	\label{sec:results}
	The filtering process discussed in Section~\ref{sec:data_assurance} yielded a set of 1734 observations (58 hours) from six pointings. Before delving into constructing the power spectrum from the observations, we compared our pipeline with the traditional MWA Real Time System pipeline \citep[RTS,][]{Mitchell2007} used in \cite{Trott2020}.
	%After acquiring the filtered set of observations, we moved to the power spectrum analysis discussed in Section~\ref{sec:power_spectrum}. We first compared our pipeline with the traditional MWA Real Time System pipeline \citep[RTS,][]{Mitchell2007} used in \cite{Trott2020}, and then generated the power spectra for the individual pointing and from the whole set of observations.
		
	\subsection{Pipeline vs $RTS$}
	\label{sec:pipeline_rts}
	
   The RTS pipeline constitutes the following: conversion of raw MWA files to uvfits using \href{https://github.com/MWATelescope/cotter}{cotter}; 2) use of AOFlagger for flagging; 3) DI calibration using a sky model ; direction dependent calibration on the five brightest sources and 4) catalogue subtraction for foreground removal. As observed, there are quite a few differences between the two pipelines.  Our pipeline uses the latest LoBES catalogue as the sky model while RTS used the catalogue created from GLEAM along with cross-matched sources from TGSS GRMT described in \citep{Procopio2017}. To reduce comparative complexities, we used the same sky model and propagated the same flags to the observation, hence the comparison here is mainly between the calibration implementation.
   
   %We used a single observation with the same sky model used in RTS to reduce comparative complexities.
   \begin{figure}
	\centering
	\includegraphics[width=0.95\linewidth]{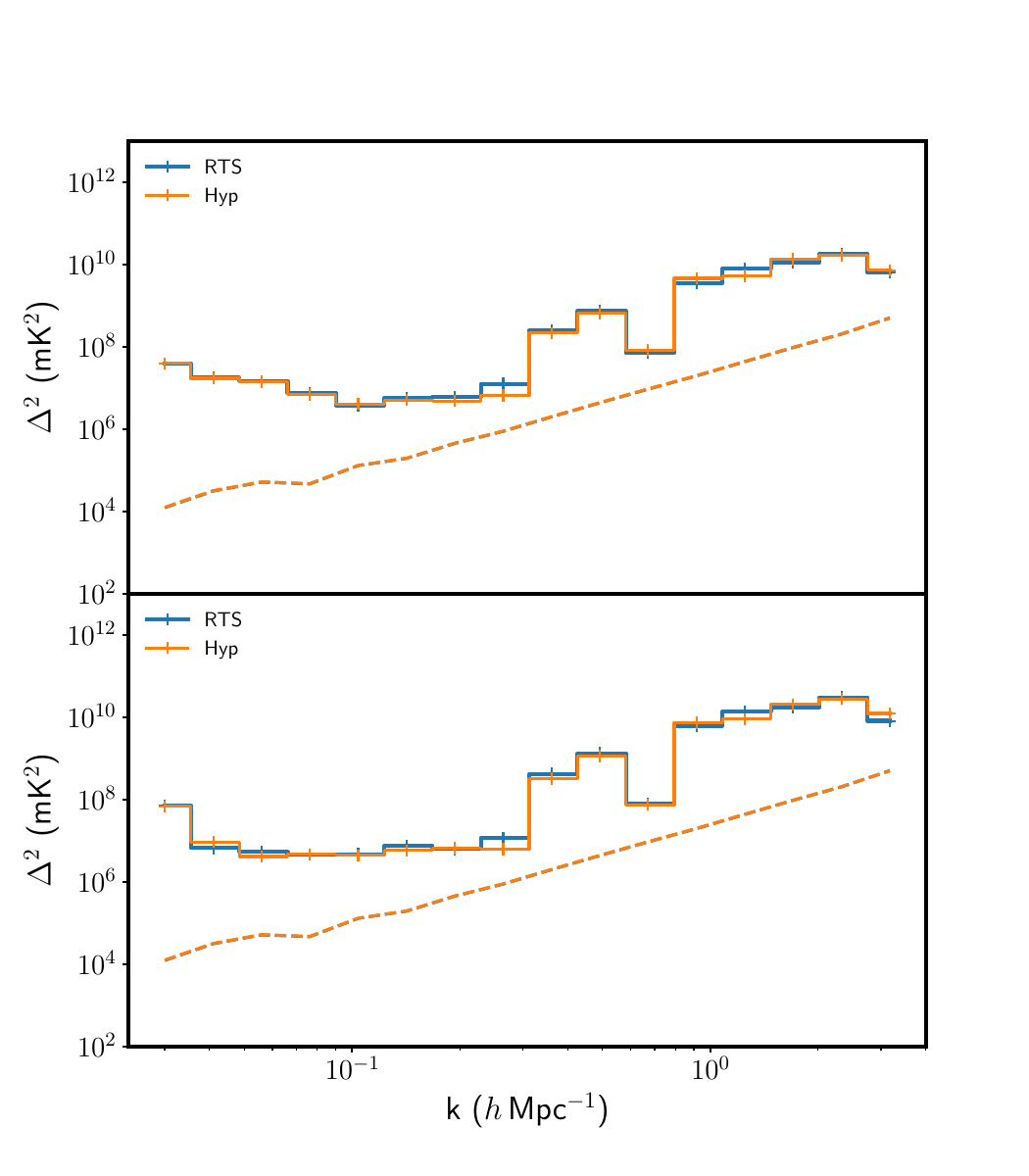}
	\caption{0ne-dimensional power spectra constructed by averaging the power over the ($k_{\perp}, k_{||}$) boundaries described in Section~\ref{sec:pipeline_rts} with top panel present results from EW polarisation and bottom panel from NS. The dotted lines is the thermal noise evaluated from the weights assigned to each of the visibility points.}
	\label{fig:pipeline_rts}
\end{figure} 
   	
   \begin{figure}
   	\centering
   	\includegraphics[width=0.95\linewidth]{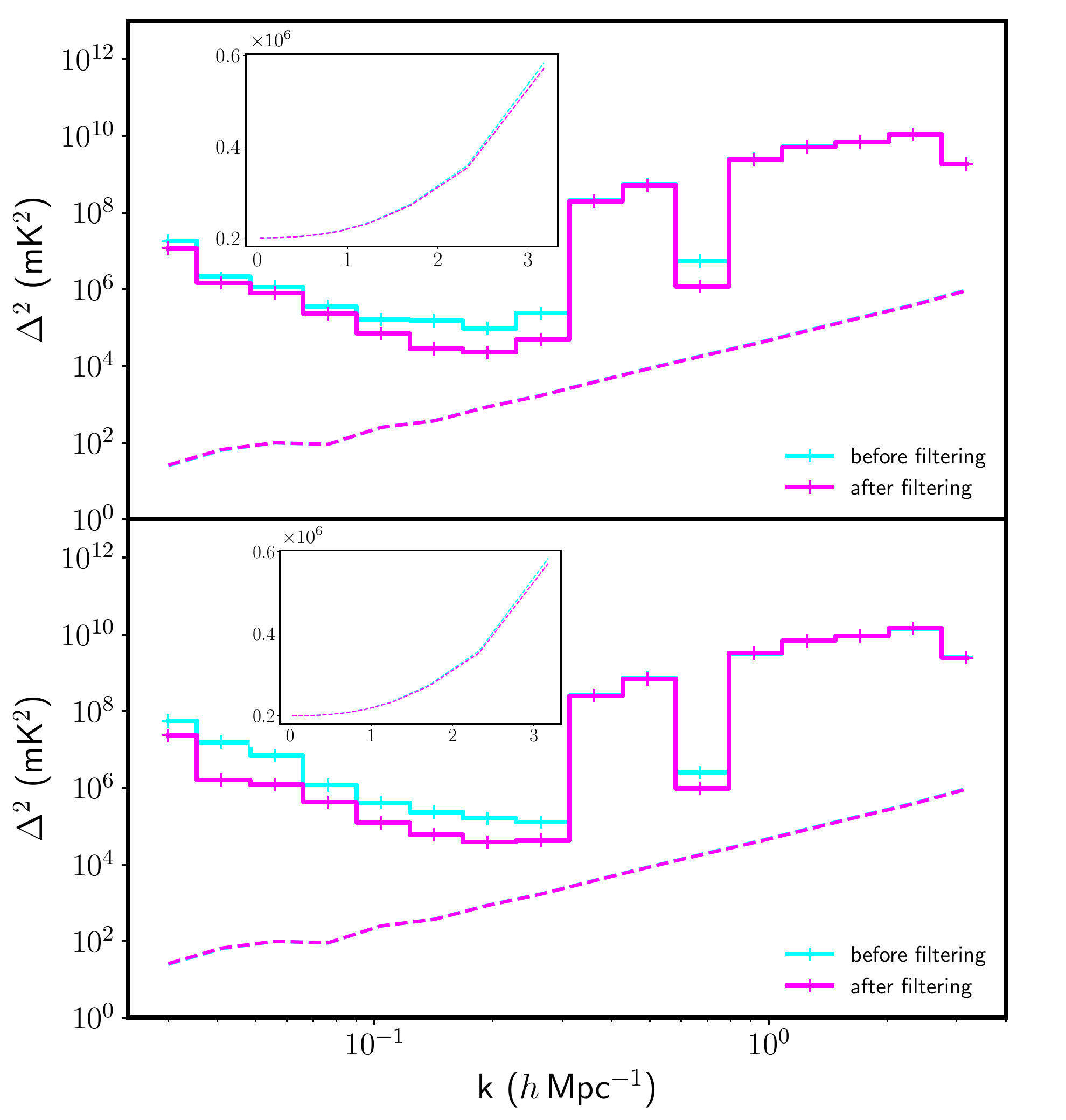}
   	\caption{0ne-dimensional power spectra constructed using 300 observations. The spectra in cyan was from observations selected from the unfiltered set of 9655 while the one in magenta was form from the filtered set of  1734 observations. The power is averaged over the ($k_{\perp}, k_{||}$) boundaries described in Section~\ref{sec:pipeline_rts} with top panel present results from EW polarisation and bottom panel from NS. The dotted lines is the thermal noise evaluated from the weights assigned to each of the visibility points. The insets at the bottom shows the non-log plot of the thermal noise where the slight difference is visible.}
   	\label{fig:pspec_improvement}
   \end{figure} 
  
   We computed the power spectrum for a single observation from both pipelines.
   Figure~\ref{fig:pipeline_rts} presents the spherically averaged power spectra resulting from both pipelines. The $k$ bins that went into the averaging are $k_{\perp} < 0.04 h\,$Mpc$^{-1}$ to exclude low contaminated modes and $k_{||} > 0.15\, h\,$Mpc$^{-1}$ to exclude any possible left over leakage. The spectra was generated across the whole frequency band (167--197~MHz). Both strategies perform similarly, with hyperdrive performing slightly better at some $k$ modes. This result is useful to us as it informs us that any improvements to the power spectrum would be primarily attributed to the data quality assurance strategy.
   
	\subsection{Improvements to the power spectrum}
	\label{sec:pspec_improvements}
	We now compare the power spectrum formed before implementing our data quality assurance strategy described in Section~\ref{sec:methodology} to our current pipeline. The first set of observations includes 300 observations randomly picked from the 9655 datasets we started off with while the second set includes 300 observations chosen from the filtered set. All pointings are included. The comparison of the one-dimensional power spectra averaged over the $k$ bins stated in Section~\ref{sec:pipeline_rts} is shown in Figure~\ref{fig:pspec_improvement}.  Our filtering strategy does show an improvement in the power level particularly at low $k$ modes indicating that excluding unreliable observations helps in preventing power to leak beyond the wedge region. This behaviour is observed along both polarisations.
	
	%However, we are still systematic dominated at high $k$ modes and we are unsure of the potential causes. We would need further diagnosis to study these systematics to mitigate them.

	\subsection{Power Spectra for each pointing}
	\label{sec:pspectra_pointing}
	We also compared the power spectra generated under the same conditions as mentioned in Section~\ref{sec:pipeline_rts} for different pointings. After the filtering process discussed in Section~\ref{sec:data_assurance}, we were left with observations for only six pointings  (3, 2, 1, 0, -1, -2). Each of these pointings carry different number of observations, therefore to avoid noise bias in our results, we used the same number of observations, amounting to about 4.3~hours.
	
\begin{figure*}
		\centering
		\includegraphics[width=\linewidth]{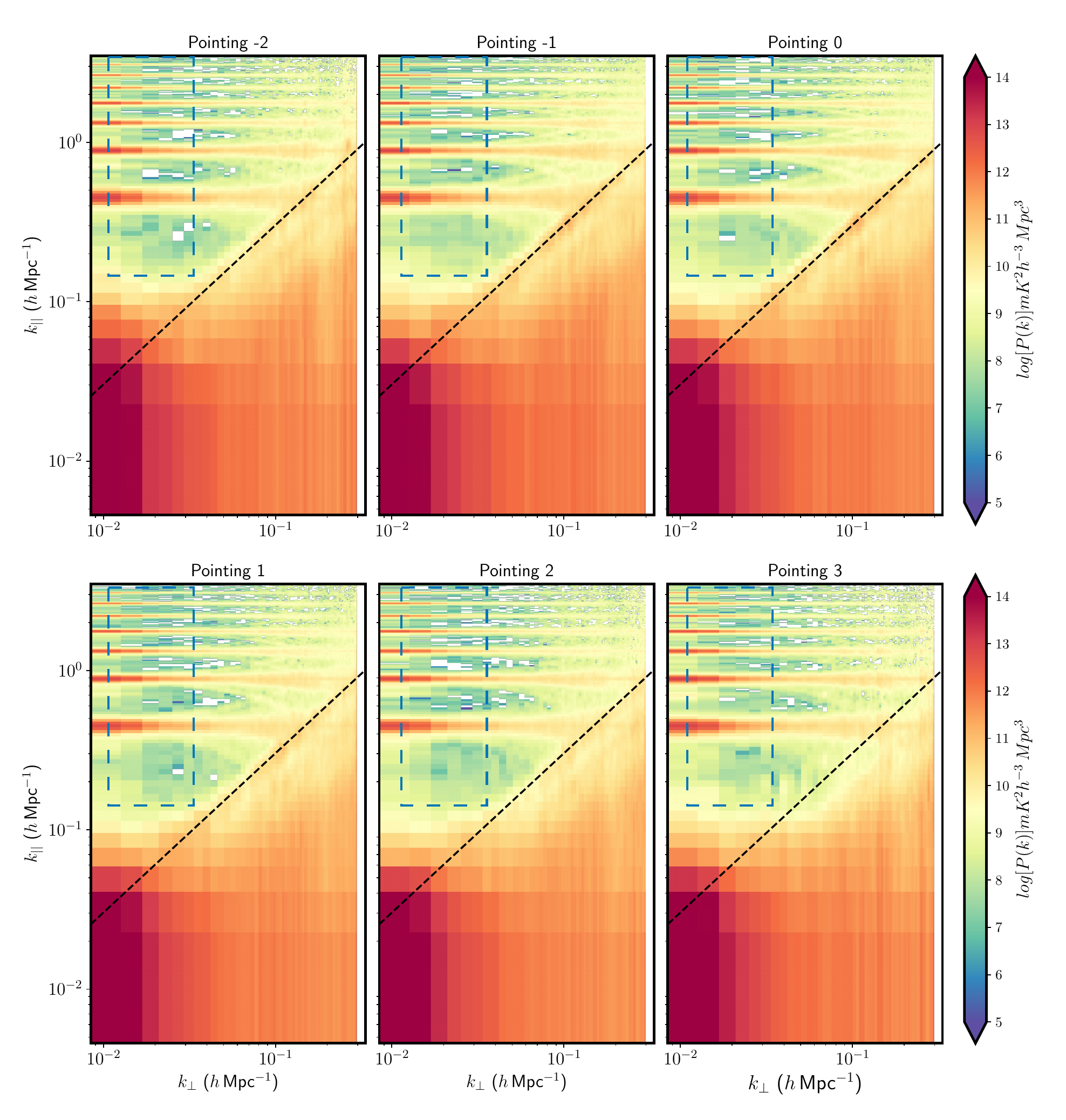}
		\caption{Two-dimensional power spectra produced from 130 observations (4.3~hours) across full frequency band for the different pointing. The power at large angular scales decrease as we shift from negative to positive pointings, showing the effect of the contribution from the Galactic plane. The dashed black lines represents the horizon limit. The dashed blue box indicates the area enclosed by $0.01 h,$Mpc$^{-1}$ $<k_{\perp} <0.04 h\, $Mpc$^{-1}$ and Mpc$^{-1} 0.15 h\,< k_{||}  < 3.4 h\, $Mpc$^{-1}$.}
		\label{fig:psectra2d_pointing}
	\end{figure*}

	The two-dimensional power spectra for EW polarisation constructed across the full band for the six pointings are displayed in Figure~\ref{fig:psectra2d_pointing}. The black dotted line marks the horizon limit marking the baseline-dependent boundaries.  The harmonics are due to the flags applied on the edges and centre frequency channels. The foreground subtraction performed a decent job in reducing the overall foreground power. 
	%particularly at large $k-$modes. 
	The bright foreground emission constrained at low $k-$modes are mostly from the Galactic emission that were not included in the subtraction model. 
	
	The EoR window we are interested in lies above the black dotted lines with a buffer of $0.05 h\, $Mpc$^{-1}$, bounded by $0.01 h\,$Mpc$^{-1}$ $<k_{\perp} <0.04 h\, $Mpc$^{-1}$ and 
	Mpc$^{-1} 0.15 h\,< k_{||}  < 3.4 h\, $Mpc$^{-1}$, region enclosed by the blue dashed lines in Figure~\ref{fig:psectra2d_pointing}. We chose these $k$ modes to escape potential foreground spilling into the window. %targetting modes close to the horizon.
	It is hard to provide a quantitative difference between the pointings as they all are behaving at different $k$ modes. The most obvious pattern, analogous to top panel of Figure~\ref{fig:image_qa}, are the modes corresponding to $0.01 h,$Mpc$^{-1}$ $< k_{\perp} < 0.015 h\, $Mpc$^{-1}$. %The power decreases from pointing $-2$ to $0$, and rises fro pointing $2$ and $3$.
 The cleanest window within the stated boundary is produced by pointings $0$, $1$ and $2$. Power leakage beyond the horizon limit is more prominent in pointings -2 and -1. Even though the Galactic plane in these pointing is past the second sidelobe of the primary beam response of MWA, it still impacts the foregrounds via aliasing as it sets over the horizon \citep{Barry2024}.
	
	%The cleanest window are produced by positive pointings. Power leakage beyond the horizon limit is more prominent in pointings -2 and -1. Even though the Galactic plane in these pointing is past the second sidelobe of the primary beam response of MWA, it still impacts the foregrounds via aliasing as it sets over the horizon \citep{Barry2024}. This behaviour is analogous to Figure~\ref{fig:image_qa}.}
	
\begin{figure*}
		\centering
		\includegraphics[width=0.9\linewidth]{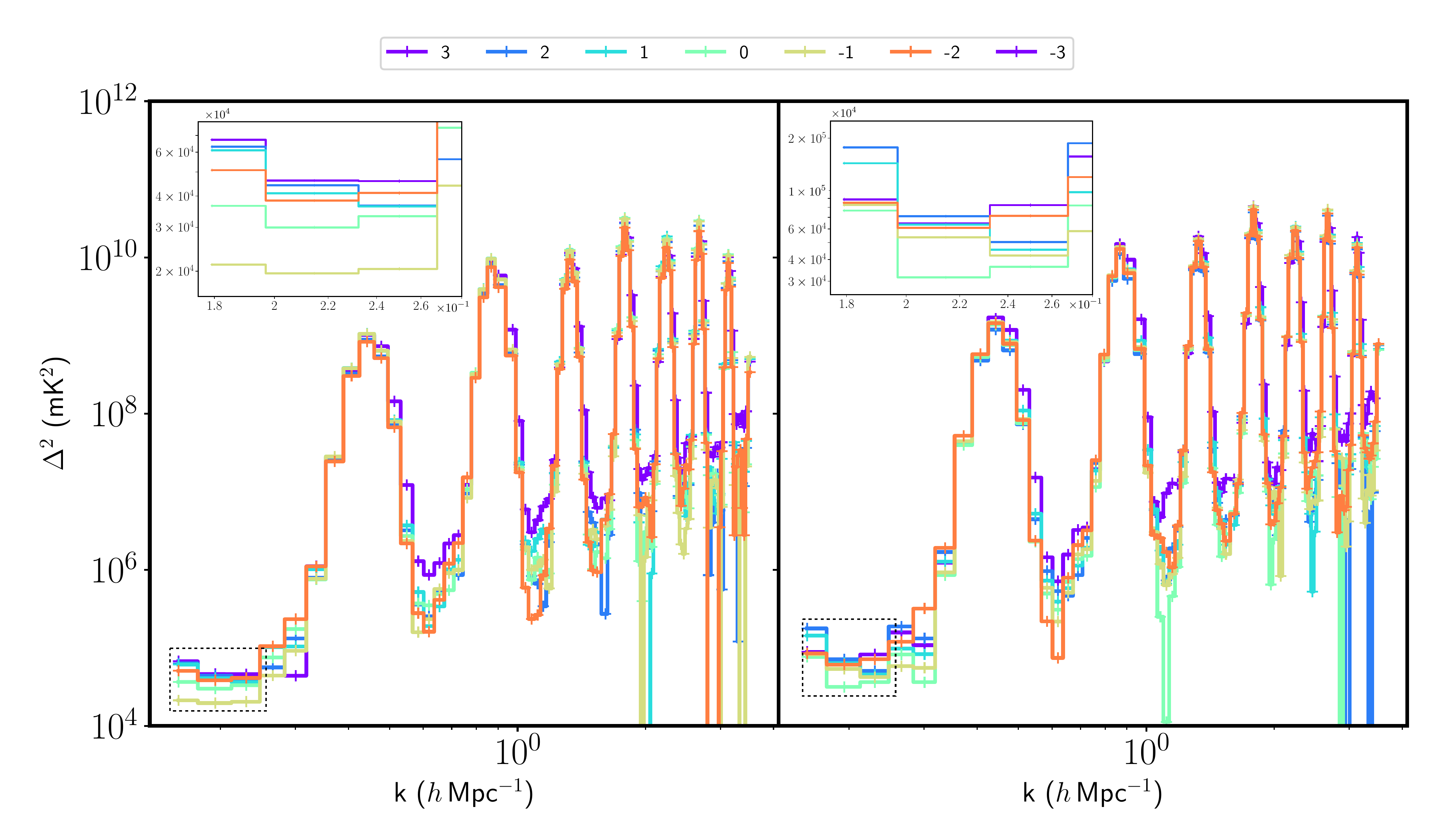}
		\caption{One-dimensional power spectra produced from 130 observations (4.3~hours) over 15~MHZ band (182 -- 197~MHz) for the different pointings, represented by solid coloured lines. The left panel shows results for EW polarisation and right panel for NS polarisation. The dotted lines show the measured thermal noise ($2\sigma$ plus sample variance) computed across the observations. The inset shows a zoom version of the distribution of the power level at low $k$ modes.}
		\label{fig:psepctra1d_pointing}
	\end{figure*}

	We also generated power spectra at $z=6.5$ using visibilities across (182, 197) MHz band. The results are similar to the full band in Figure~\ref{fig:psectra2d_pointing}. The dimensionless power spectra $\Delta^2$, averaged over $ 0.01 <k_{\perp} < 0.04 \,h$ Mpc$^{-1}$ and $k_{||} > 0.15\, h$ Mpc$^{-1}$, is plotted in Figure~\ref{fig:psepctra1d_pointing}. The positive pointings have a lower floor compared to the negative ones at large angular scales for both polarisations supporting our statement about the Galactic plane contribution across these modes. The inset shows a clear representation of the power level  for each pointing at low $k$ modes. At high $k$ modes the pointings seem to converge.
	
	Although some of the pointings performs better than other at specific $k$ modes, the distribution of the power spectra do not indicate any far-flung behaviours. Therefore, we proceeded with the power spectrum analysis using observations from all six pointings.
	
 \begin{figure}
   	\centering
    
   	\includegraphics[width=1.0\linewidth]{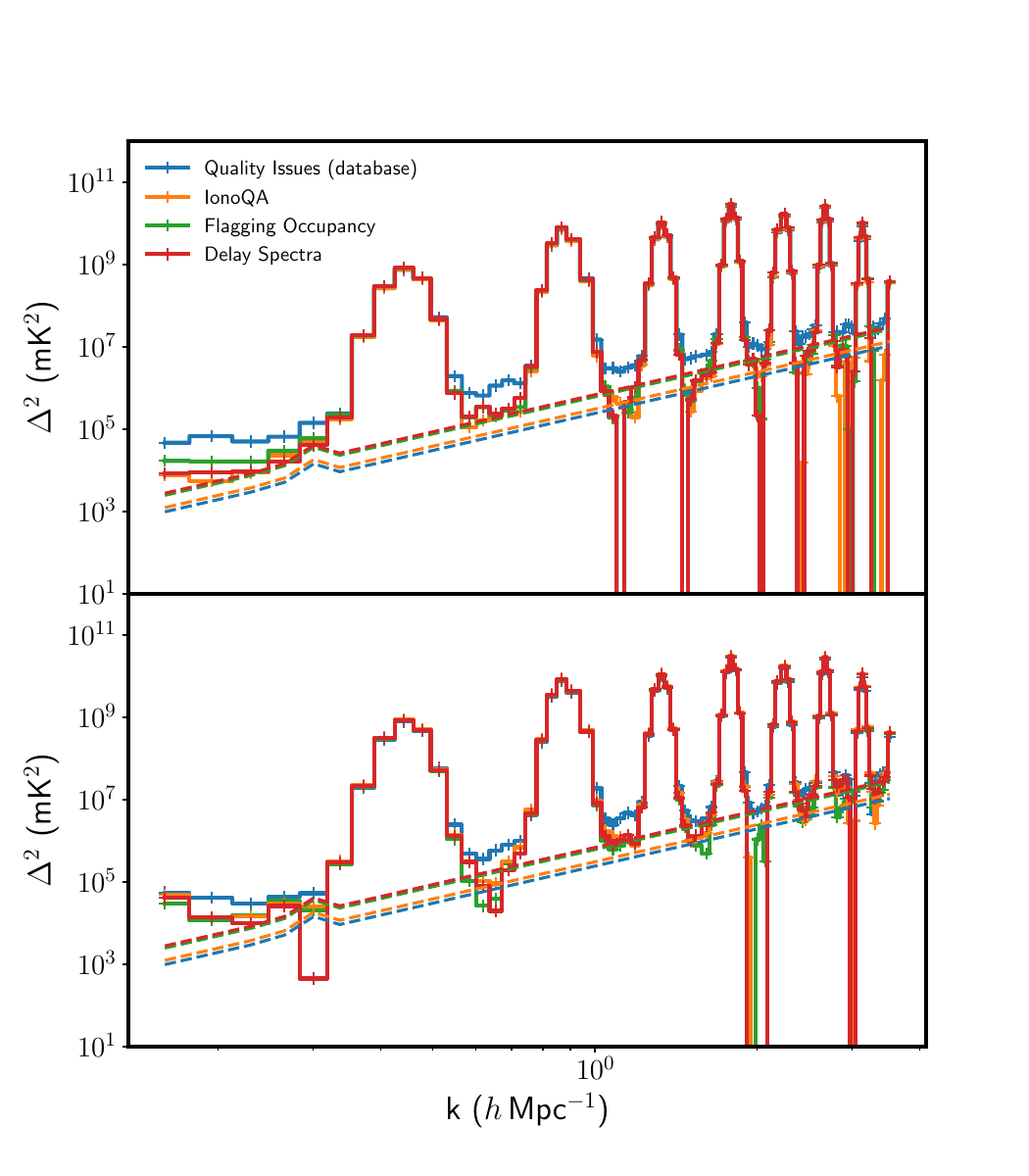}
   	\caption{One-dimensional power spectra produced over 15~MHZ band (182 -- 197~MHz) from observations as they are ruled out by the different steps namely: quality issues (300 observations), ionoQA (238 observations), flagging occupancy (130 observations) and delay spectra (100 observations). The dotted lines show the measured thermal noise ($2\sigma$ plus sample variance) computed from the observations.}
   	\label{fig:pspec_cuts_difference}
   \end{figure} 
   
		\begin{figure}[h]
		\centering
		\includegraphics[width=\linewidth]{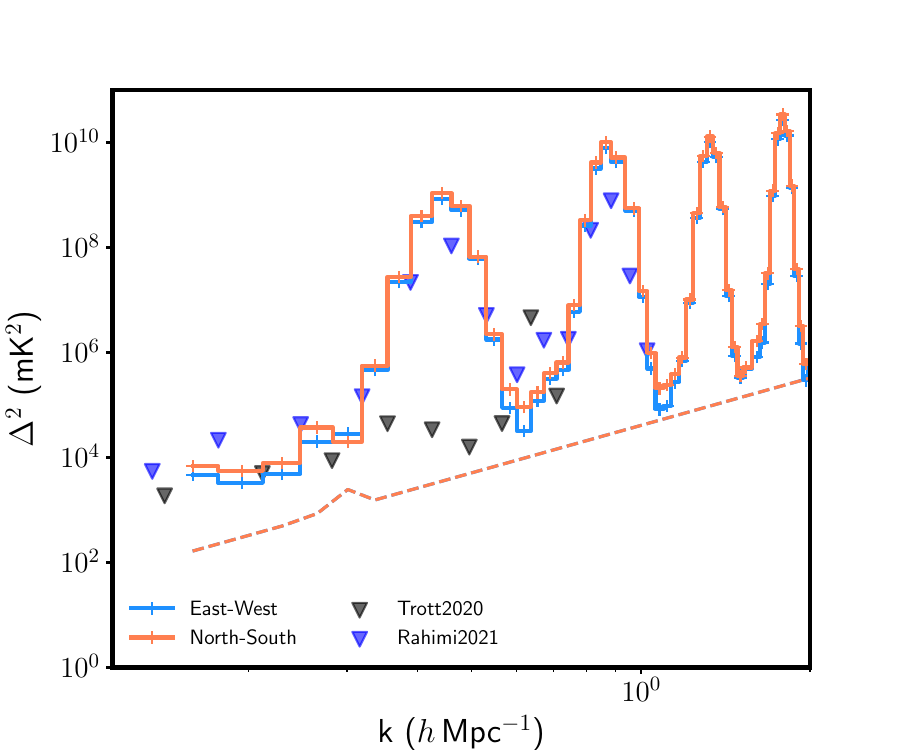}
		\caption{One-dimensional power spectra produced from 1796 observations (58~hours) over 15~MHZ band (182 -- 197~MHz) for both polarisations. The dotted lines show the measured thermal noise ($2\sigma$ plus sample variance) computed from 1734 observations. The grey and blue markers indicate the upper limits obtained in \citet{Trott2020} and \citet{Rahimi2021} respectively.}
		\label{fig:1D_pspectra_combined}
	\end{figure}

	\subsection{Validating our strategy}
	 We analyzed our systematic strategy on a set of observations across all pointings, chosen arbitrarily. Our aim was to construct and compare the power spectra at each of the filtering steps. However, we were limited by our pipeline settings and computational resources, preventing us from constructing the power spectra after data quality issues were reported in the database. Therefore, we began with a set of 300 observations that had passed data quality inspection.
 
    Out of these 300 observations, 62 were found to have an ionospheric metric greater than 5. Discarding highly-ionospherically active observations produced a difference of about an order in power at most of the k-modes, as illustrated by the one-dimensional power spectra in Figure~\ref{fig:pspec_cuts_difference}. Applying the flagging occupancy left us with 130 observations, raising the power level higher. This rise is due to the reduction in the number of observations, resulting in a higher noise level, delineated by the thermal noise in dashed lines. The metrics from the delay-transformed visibilities rejected 30 observations, slightly improving the results. The image statistics did not spot any misbehaving observations for this set.

    Since it is not straightforward to validate the power spectra due to the varying number of observations, we use the gap between the power and the thermal noise to infer any improvements. As mentioned in the previous paragraph, the power spectra evaluated after filtering out using the ionoQA show major improvements. Comparing power spectra produced after accounting for flagging occupancy and metrics evaluated from delay spectra also indicate an improvement when the contaminated observations indicated by the delay spectra metrics are removed.

    At this point, it is challenging to discuss the specific improvements of flagging occupancy over ionospheric filtering using the same principle. However, the distance of the power from the corresponding thermal noise level does exhibit an improvement. After discarding observations ruled out by flagging occupancy, the power is closer to the thermal noise compared to before filtering these observations, indicating a cleaner dataset.
    
	\subsection{Combined Power spectra}
	In the final step, we combined the filtered observations (58~hours) described in Section~\ref{sec:data_assurance} and formed power spectra from the individual polarisations for diagnosis purposes as this paper is intended to present the improved pipeline, discussing explicitly the systematic mitigation approach.
	 The resulting one-dimensional power spectra at $z=6.5$ obtained by averaging over $0.01 <k_{\perp} < 0.04 \,h$ Mpc$^{-1}$ and $k_{||} > 0.15\, h$ Mpc$^{-1}$, is shown in Figure~\ref{fig:1D_pspectra_combined}. The lowest measurements for EW and NS polarisations are $\Delta^2= (57.2 \textrm{ mK})^2$ and $\Delta^2= (74.6 \textrm{ mK})^2$ at $k=0.19\,h$ Mpc$^{-1}$ respectively.
	
	We overlaid upper limits from \citet{Trott2020} and \citet{Rahimi2021}. However, these upper limits cannot be directly compared with our measurements due to differences in calculation conditions. \citet{Rahimi2021} utilised only phase I data and focused on a different observing field. On the other hand, \citet{Trott2020} used observations from the same field but included both phase I and phase II data, employing a different sky model. Incorporating the sky model from \citet{Trott2020} yielded marginal differences. To enhance compatibility, averaging was performed on the same $k$ modes as in \citet{Trott2020}. 
 
 %Consequently, the results differ from those shown in Figure~\ref{fig:psepctra1d_pointing}, particularly at high $k$ modes, depicting higher power. This increase in power is attributed to the presence of bins contaminated by harmonics during averaging.
	
	A back-of-the-envelop comparison of the sensitivity levels between a single phase I and phase II observation yielded an improvement of about 0.6. Applying this factor to our current measurements does indicate an improvement to our current results which is promising.
	
	\section{Conclusion and Future Work}
	\label{sec:conclusion}
	In this paper, we presented a statistical framework whereby metrics were derived at intermediate stages to prevent systematic errors from propagating to the power spectrum. These metrics were used to assess the quality of an observation, informing the pipeline whether it should proceed to the next stage. We found that without these metrics, many systematics, such as bad frequency channels, malfunctioning antennas, and corrupted observations, would not have been identified. When compared to observations from the EoR0 field used in the estimation by \citep{Trott2020}, about one third of the observations in common were flagged by our methodology.
	
	Our strategy filtered out 82\% of our initial observations, leaving approximately 58 hours of data (half the number used in \citep{Trott2020}). We achieved a comparable lowest floor of $\Delta^2= (57.2 \textrm{ mK})^2$ at $k=0.19,h$ Mpc$^{-1}$ at $z=6.5$ along the EW polarisation. These results were obtained from observations less dominated by systematic errors, as determined by our statistical framework, increasing the reliability and confidence in our power spectrum results.
	
	We also evaluated the accuracy and reliability of the calibration software Hyperdrive \citep{Jordan}. Furthermore, this work explores the latest sky model LoBES, designed specifically for EoR experiments targeting the EoR0 field. Overall, our current processing pipeline implemented in NextFlow \citep{Di2023} is efficient, with most integrated software components working harmoniously with minimal human intervention, thereby reducing errors.

The methodology and results presented in this paper can be improved upon and we have identified few avenues:
	\begin{enumerate}
		\item Including observations from the Phase II compact configuration would help in obtaining lower power levels as demonstrated by our rough estimations.
		\item Increasing the number of iterations for direction-independent calibration. This would enhance convergence of the algorithm producing more accurate complex antenna gains.
		\item Improving on the current calibration algorithm using gain solutions from the autocorrelations \citep{Li2019}.
		\item Strengthening our current data quality framework, by automating the anomaly detection from the phases of the complex antenna gains and incorporating machine learning.
		\item Clustering observations sharing similar features using the derived statistical metrics to identify the optimal cluster of observations for power spectrum estimation.
\end{enumerate}
Some of the aforementioned strategies are already in development and will be featured in \citep{Nunhokee}.

	\section{Acknowledgements}
	This research was supported
	by the Australian Research Council Centre of Excellence for
	All Sky Astrophysics in 3 Dimensions (ASTRO 3D), through
	project number CE170100013. 
	%The International Centre forRadio Astronomy Research (ICRAR) is a Joint Venture ofCurtin University and The University of Western Australia, funded by the Western Australian State government. 
	This scientific work uses data obtained from \textit{Inyarrimanha Ilgari Bundara} / the Murchison Radio-astronomy Observatory. We acknowledge the Wajarri Yamaji People as the Traditional Owners and native title holders of the Observatory site. Establishment of CSIRO's Murchison Radio-astronomy Observatory is an initiative of the Australian Government, with support from the Government of Western Australia and the Science and Industry Endowment Fund. Support for the operation of the MWA is provided by the Australian Government (NCRIS), under a contract to Curtin University administered by Astronomy Australia Limited. This work was supported by resources provided by the Pawsey Supercomputing Research Centre with funding from the Australian Government and the Government of Western Australia. The International Centre for Radio Astronomy Research (ICRAR) is a Joint Venture of Curtin University and The University of Western Australia, funded by the Western Australian State government.  Some of these data and the pipeline development was undertaken with industry partners, Downunder Geosolutions, and we acknowledge their role in the progress of this project. We thank Michael Wilensky for his useful suggestions on the paper draft. N.B. acknowledges the support of the Forrest Research Foundation, under a postdoctoral research fellowship.

	\section{Data Availability}
	The data used in this paper is publicly available on MWA Archival System.

	\printbibliography
%\bibliography{bibliography}
\end{document}